\documentclass[aps,prb,twocolumn,superscriptaddress]{revtex4-2}

\usepackage{amsmath}
\usepackage{float}
\usepackage{graphicx}
\usepackage{siunitx}
\usepackage{amsfonts}
\usepackage{multirow}
\usepackage{amssymb}
\usepackage{soul}
\usepackage[table]{xcolor}
\usepackage{hhline}
\usepackage{comment}
\usepackage{array}
\usepackage[colorlinks=true]{hyperref}

\newcommand\Tstrut{\rule{0pt}{2.2ex}}
\newcommand\Bstrut{\rule[-1ex]{0pt}{0pt}}

\begin{document}

    \title{Low-energy modeling of three-dimensional topological insulator nanostructures}
    \author{Edu\'ard Zsurka}
    \thanks{eduard.zsurka@uni.lu}    
    \affiliation{Peter Grünberg Institute (PGI-9), Forschungszentrum Jülich, 52425 Jülich, Germany}
    \affiliation{JARA-Fundamentals of Future Information Technology, Jülich-Aachen Research Alliance, Forschungszentrum Jülich and RWTH Aachen University, 52425 Jülich, Germany}
    \affiliation{Department of Physics and Materials Science, University of Luxembourg, 1511 Luxembourg, Luxembourg}
    \author{Cheng Wang}
    \affiliation{Peter Grünberg Institute (PGI-1), Forschungszentrum Jülich and JARA, 52425 Jülich, Germany}
    \author{Julian Legendre}
    \affiliation{Department of Physics and Materials Science, University of Luxembourg, 1511 Luxembourg, Luxembourg}
    \author{Daniele Di Miceli}
    \affiliation{Department of Physics and Materials Science, University of Luxembourg, 1511 Luxembourg, Luxembourg}
    \affiliation{Institute for Cross-Disciplinary Physics and Complex Systems IFISC (CSIC-UIB), E-07122 Palma, Spain}
    \author{Llorenç Serra}
    \affiliation{Institute for Cross-Disciplinary Physics and Complex Systems IFISC (CSIC-UIB), E-07122 Palma, Spain}
    \affiliation{Department of Physics, University of the Balearic Islands, E-07122 Palma, Spain}
    \author{Detlev Gr\"utzmacher}
    \affiliation{Peter Grünberg Institute (PGI-9), Forschungszentrum Jülich, 52425 Jülich, Germany}
    \affiliation{JARA-Fundamentals of Future Information Technology, Jülich-Aachen Research Alliance, Forschungszentrum Jülich and RWTH Aachen University, 52425 Jülich, Germany}
    \author{Thomas L. Schmidt}
    \affiliation{Department of Physics and Materials Science, University of Luxembourg, 1511 Luxembourg, Luxembourg}
    \author{Philipp Rüßmann}
    \affiliation{Peter Grünberg Institute (PGI-1), Forschungszentrum Jülich and JARA, 52425 Jülich, Germany}
    \affiliation{Institute for Theoretical Physics and Astrophysics, University of Würzburg, 97074 Würzburg, Germany}
    \author{Kristof Moors}
    \thanks{k.moors@fz-juelich.de}
    \affiliation{Peter Grünberg Institute (PGI-9), Forschungszentrum Jülich, 52425 Jülich, Germany}
    \affiliation{JARA-Fundamentals of Future Information Technology, Jülich-Aachen Research Alliance, Forschungszentrum Jülich and RWTH Aachen University, 52425 Jülich, Germany}
	
\date{\today}
	
\begin{abstract}
We develop an accurate nanoelectronic modeling approach for realistic three-dimensional topological insulator nanostructures and investigate their low-energy surface-state spectrum. Starting from the commonly considered four-band $\boldsymbol{\mathrm{k\cdot p}}$ bulk model Hamiltonian for the Bi$_2$Se$_3$ family of topological insulators, we derive new parameter sets for Bi$_2$Se$_3$, Bi$_2$Te$_3$ and Sb$_2$Te$_3$. 
We consider a fitting strategy applied to \emph{ab initio} band structures around the $\Gamma$ point that ensures a quantitatively accurate description of the low-energy bulk and surface states, while avoiding the appearance of unphysical low-energy states at higher momenta, something that is not guaranteed by the commonly considered perturbative approach. 
We analyze the effects that arise in the low-energy spectrum of topological surface states due to band anisotropy and electron-hole asymmetry, yielding Dirac surface states that naturally localize on different side facets. 
In the thin-film limit, when surface states hybridize through the bulk, we resort to a thin-film model and derive thickness-dependent model parameters from \emph{ab initio} calculations that show good agreement with experimentally resolved band structures, unlike the bulk model that neglects relevant many-body effects in this regime. Our versatile modeling approach offers a reliable starting point for accurate simulations of realistic topological material-based nanoelectronic devices.

\end{abstract}

\maketitle

\section{Introduction}
\label{sec:introduction}

Topological insulators (TIs) are a novel class of materials that have garnered substantial interest in the recent decades due to their possible application in electronics, spintronics and quantum information processing~\cite{hasan_2011, Breunig2022}. TIs are characterized by the existence of topologically protected states at the boundaries of a sample, which are protected against any local perturbations that respect time-reversal symmetry. The Bi$_2$Se$_3$ family of materials, here referring to Bi$_2$Se$_3$, Bi$_2$Te$_3$ and Sb$_2$Te$_3$, are three-dimensional (3D) time-reversal-invariant TIs with a large inverted gap. They have a layered structure, consisting of five-atom, or quintuple layers (QL) arranged along the $\hat{z}$ direction (see Fig.~\ref{fig:spec}a). The bulk electronic structure is described by a nontrivial $Z_2$ topological invariant, which ensures the existence of protected spin-nondegenerate surface states with massless Dirac-like dispersion.

The Bi$_2$Se$_3$ family of 3D TIs have been first described in 2009, when their topological properties were uncovered and a four-band $\boldsymbol{\mathrm{k\cdot p}}$ Hamiltonian describing the bulk dispersion was proposed~\cite{Zhang2009}. Material parameters of this Hamiltonian were obtained using perturbation theory~\cite{Liu2010B, nechaev2016}, which yields an accurate description of the electronic band structure at $\Gamma$ at low energies. However, the degree to which the vicinity of $\Gamma$ is captured by the obtained parameters varies from case to case. When used for nanoelectronic device simulations, the model has to accurately capture the entire region in momentum space over which the Dirac cone of the topological surface states stretches out, while also remaining well behaved at larger momenta such that unphysical electronic states do not appear at low energies. This is not always guaranteed by applying perturbation theory at $\Gamma$, which motivates us to derive a new set of parameters that can give a good quantitative description of the Bi$_2$Se$_3$ family of materials. In this work, we obtain the parameters of the four-band $\boldsymbol{\mathrm{k\cdot p}}$ Hamiltonian for Bi$_2$Se$_3$, Bi$_2$Te$_3$ and Sb$_2$Te$_3$ by an alternative method. We fit the model to \emph{ab initio} band structures in such a way that the vicinity of $\Gamma$ is accurately considered up to sufficiently large momenta and all the relevant features of the band structure (e.g., the topology) are taken into account.

With the newly obtained material parameters, we analyze the low-energy spectrum of experimentally relevant nanostructures. We consider the effects of band structure anisotropy and electron-hole asymmetry, which modify the dispersion of the surfaces states on surfaces with different orientations.
While usually ignored for resolving the transport properties related to Dirac surface states in nanostructure systems~\cite{Bardarson2010, brey2014, Bardarson_2018, moors2018, ziegler_2018}, we find that the low-energy spectrum can be significantly affected by this anisotropy and asymmetry for some of the materials under consideration here.

\begin{table*}[t]
    \centering
    \begin{tabular}{cl|cccc|ccc|ccc}
    \toprule
    && Bi$_2$Se$_3$~\cite{Zhang2009}&Bi$_2$Se$_3$~\cite{Liu2010B}&Bi$_2$Se$_3$~\cite{nechaev2016}&Bi$_2$Se$_3$ fit&Bi$_2$Te$_3$~\cite{Liu2010B}&Bi$_2$Te$_3$~\cite{nechaev2016}&Bi$_2$Te$_3$ fit&Sb$_2$Te$_3$~\cite{Liu2010B}&Sb$_2$Te$_3$~\cite{nechaev2016}&Sb$_2$Te$_3$ fit\Tstrut\Bstrut\\
    \hline         
     $A_0$&$[\text{eV\AA}]$      & 4.1 &   3.33 &   2.51 &   4.33 &   2.87 &    4 &   4.40 &   3.4 &   3.7 &   3.89 \Tstrut \\
     $B_0$&$[\text{eV\AA}]$    & 2.2   &   2.26 &   1.83 &   1.94 &   0.3 &    0.9 &   0.55 &   0.84 &   1.17 &   1.69 \\
     $C_0$&$[\text{eV}]$                &-0.0068 &  -0.0083 &   0.048 & -0.28 &  -0.18 &   -0.12 &  -0.014 &   0.001 &   0.02 & 0.10 \\
     $C_1$&$[\text{eV\AA}^2]$ & 1.3    &   5.74 &   1.41 &  1.46 &   6.55 &    2.67 &  1.65 & -12.39 & -14.2 &  -6.48 \\
     $C_2$&$[\text{eV\AA}^2]$ &19.6    &  30.4  &  13.9 &   22.81 &  49.68 &  154.35 &  29.47 & -10.78 &  -6.97 &  -4.26 \\
     $M_0$&$[\text{eV}]$     &-0.28   &  -0.28 &  -0.17 &  -0.30 &  -0.3 &   -0.3 &  -0.26 &  -0.22 &  -0.18 &  -0.21 \\
     $M_1$&$[\text{eV\AA}^2]$ &10      &   6.86 &   3.35 &   6.00 &   2.79 &    9.25 &   4.62 &  19.64 &  22.12 &  19.32 \\
     $M_2$&$[\text{eV\AA}^2]$ &56.6    &  44.5 &  29.35 &  70.38 &  57.38 &  177.23 &  72.80 &  48.51 &  51.28 &  63.91\Bstrut \\
     \hline
     Bulk gap & [meV] & 560 & 344 & 280 & 472 & - & 155 & 303 & 135 & 155 & 303 \Tstrut \\
     $\lambda_z$&$[\text{\AA}]$& 9.01 & 12.83 & 9.87 & 6.0 & - & 19.68 & 15.8 & 36.28 & 28.99 & 21.48 \\
     $\zeta_\text{DP}$& & 0.57 & 0.92 & 0.71 & 0.62 & 1.67 & 0.64 & 0.67 & 0.18 & 0.18 & 0.31\\
     $\Delta E_{\text{DP}}$ &$[\text{meV}]$&61&43&9&24&445&175&12& 90 & 91 & 56\Bstrut\\
     \toprule
    \end{tabular}
    \caption{Parameters for the $\boldsymbol{\mathrm{k\cdot p}}$ bulk model Hamiltonian \eqref{eq:H_3D} for the three studied materials: Bi$_2$Se$_3$, Bi$_2$Te$_3$ and Sb$_2$Te$_3$. The parameters taken from Refs.~\cite{Liu2010B, nechaev2016} are obtained through perturbation theory applied at $\Gamma$. We introduce a new set of parameters obtained by fitting to \emph{ab initio} band structure data around $\Gamma$. Additionally, the bulk gap, the penetration depth $\lambda_z$ along $z$ of the surfaces states in the $x\text{-}y$ plane, the relative position $\zeta_\text{DP}$ of the Dirac point of a slab parallel to the $x\text{-}y$ plane with respect to the bulk gap at $\Gamma$ (see Eq.~\eqref{eq:zeta_DP}), and the energy difference $\Delta E_{\text{DP}}$ between the Dirac points of surfaces states parallel to the $x\text{-}y$ and $x\text{-}z$ planes, are also included (see Sec.~\ref{sec:models} for details).}
    \label{tab:param_fit}
\end{table*}

When the thickness of a nanoribbon approaches a few QLs, the surface states localized on top and bottom surfaces can hybridize, leading to a gap opening in the surface-state dispersion at $\Gamma$~\cite{Linder2009, Liu2010A, Shan_2010, Lu2010}. In this case, we refer to the system as being in the \emph{thin-film limit}.
If the hybridization of the surface states is accompanied by an inversion of the surface-state spectrum, the system enters a quantum spin Hall insulator (QSHI) regime. Such thin-film geometries have attracted significant interest, being suitable for studying QSHI edge channels~\cite{Zhang_2015, Leis_2022}, the quantum anomalous Hall effect~\cite{Yu2010,Chang2013, Kou2013,Wang_2015,Yasuda2017,Sun_2019,Tokura2019,Yasuda2020,Wang2021,Qiu_2022,Atanov2024}, and topological superconductivity~\cite{Wang2015_chiral,Zeng2018,Chen2018, Mandal2022, uday2023induced}.

According to early theoretical work on 3D TI thin-films, the gap at $\Gamma$, which here we refer to as the hybridization gap $\Delta E_\text{hyb}$, shows an oscillatory behavior between a QSHI and a normal insulator (NI) state when the thickness of the thin-film is varied~\cite{Linder2009, Liu2010A, Lu2010}. However, more recent results suggest that many-body effects arising in the thin-film limit modify the oscillations and the size of the hybridization gap, giving a better agreement with experimental results~\cite{Forster2015,Forster2016,Wang2019,Liu2023}. To describe this limit, we employ an effective thin-film model that captures solely the surface-state dispersion. We extract the material parameters of the thin-film model by fitting to the surface-state spectra obtained from \emph{GW} calculations of thin films, which take into account the relevant many-body effects, for thicknesses ranging from 2 to 6 QL.

This paper is structured as follows. In Sec.~\ref{sec:models}, we give an overview of the models used to describe 3D TIs in the bulk, at the surface, and in the thin-film limit, and also discuss the material parameters. In Sec.~\ref{sec:nano}, we present the dispersion of quasi-one-dimensional nanostructures and analyze the effect of anisotropy and electron-hole asymmetry. In Sec.~\ref{sec:thin-film}, we treat the thin-film limit using the bulk model and compare the results to experimental findings. We also provide thickness-dependent material parameters for the effective thin-film model. Finally, in Sec.~\ref{sec:disc}, we interpret our findings and discuss other aspects that may be relevant for accurate nanoelectronic device modeling with 3D TIs.

\section{Models}
\label{sec:models}

\subsection{Bulk model}
The low-energy bulk electronic structure of the Bi$_2$Se$_3$ family of materials around $\Gamma$ ($\boldsymbol{k}=\boldsymbol{0}$) can be described using a four-band model, where only the valence and conduction bands that are responsible for the band inversion are considered. The bulk model Hamiltonian can be written in the following form~\cite{Zhang2009,Liu2010B},
\begin{eqnarray}
H_\text{bulk}(\boldsymbol{k})&=&\epsilon(\boldsymbol{k}) + \mathcal{M}(\boldsymbol{k})\tau_z \nonumber \\ 
&&+\,A_0(k_y\sigma_x-k_x\sigma_y)\tau_x + B_0 k_z \tau_y, \label{eq:H_3D} \\
\epsilon(\boldsymbol{k}) &=& C_0 + C_1 k_z^2 + C_2\big(k_x^2+k_y^2\big),\nonumber\\
\mathcal{M}(\boldsymbol{k}) &=& M_0+M_1 k_z^2+M_2\big(k_x^2+k_y^2\big),\nonumber
\end{eqnarray}
with $\sigma_i$, $\tau_i$ ($i = x,y,z$) the Pauli matrices for the spin and orbital degree of freedom and model parameters $A_0,B_0,C_0,C_1,C_2,M_0,M_1,M_2\in \mathbb{R}$ that can be obtained from \emph{ab initio} calculations or perturbation theory. This bulk model describes an insulator only when $|C_1|<|M_1|$ and $|C_2|<|M_2|$, avoiding the closing of the band gap at large values of $|\boldsymbol{k}|$~\cite{shen_book}, which is necessary to avoid the appearance of unphysical states at low energies upon applying the finite-difference method. The band inversion and consequently the topological properties are determined by $M_0,M_1$ and $M_2$ -- only when $M_0M_1<0$ and $M_0M_2<0$ are satisfied, Eq.~\eqref{eq:H_3D} describes a topologically nontrivial system~\cite{Liu2010B}. $C_0, C_1$ and $C_2$ are responsible for the electron-hole asymmetry, while $A_0$ and $B_0$ can be interpreted as the group velocities (up to a factor of $\hbar$) of the surface states on surfaces orthogonal to any in-plane ($x$-$y$) direction, and to the $\hat{z}$ direction, respectively. The anisotropy of the band structure is captured by different values for the corresponding in-plane and out-of-plane terms $A_0\neq B_0$, $C_{1}\neq C_{2}$ or $M_{1}\neq M_{2}$. Finite values of $M_1$ and $M_2$ prevent the fermion doubling problem, hence the Hamiltonian~\eqref{eq:H_3D} can be safely discretized on a lattice without acquiring unphysical Dirac points (at e.g. $k_{x,y,z}=\pm\pi/a$ if one considers a cubic lattice with lattice constant $a$~\cite{NIELSEN198120}), making the model suitable for modeling the low-energy spectrum of TI nanostructures with arbitrary shapes~\cite{moors2018}.

\subsection{Material parameters}

\begin{figure*}[t]
    \centering
    \includegraphics[width=\linewidth]{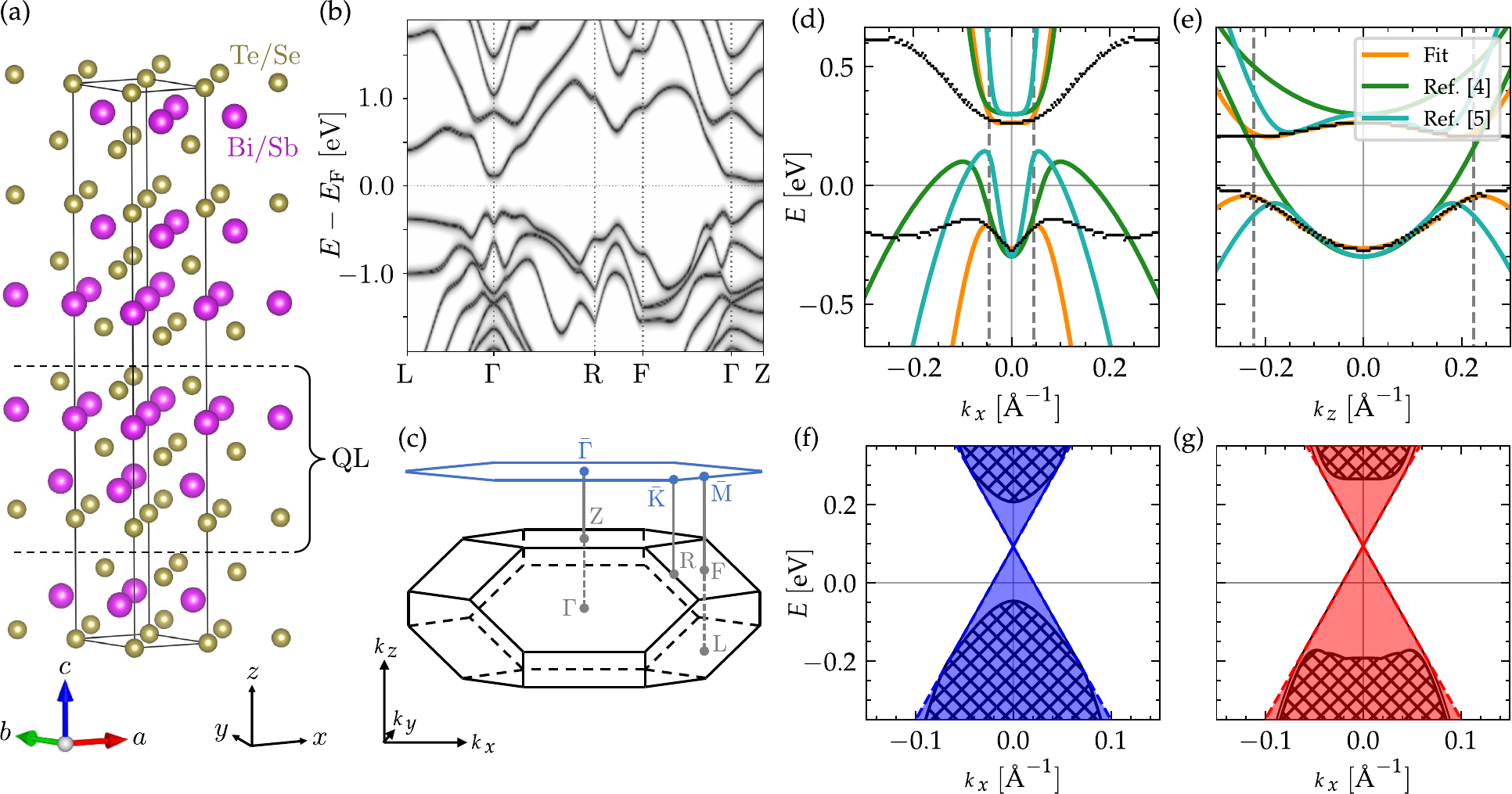}
    \caption{(a) Crystal structure of the Bi$_2$Se$_3$ family of materials. (b) Band structure for Bi$_2$Te$_3$. (c) Brillouin zone with space group $R\bar{3}m$. The blue hexagon is the 2D Brillouin zone of the projected (1,1,1) surface, and the high-symmetry points $\Gamma$, K and M are labeled. The dispersion along (d) $k_x$ and (e) $k_z$ of the bulk model (orange) that is fitted to the conduction and valence band of Bi$_2$Se$_3$ around the $\Gamma$ point, obtained from \emph{ab initio} calculations (black dots), compared to the dispersion evaluated with the material parameters taken from Ref.~\cite{Liu2010B} (green) and Ref.~\cite{nechaev2016} (cyan). The upper limits of the fit in momentum space $k_{\parallel,z}^{\text{max}}$ (see Appendix~\ref{app:fit}), are shown with gray dashed lines. Dispersion in $k_x$ of the surface state of Bi$_2$Te$_3$, with the surface in the (f) $x\text{-}y$ and (g) $x\text{-}z$ plane, evaluated analytically (dashed lines) and numerically (solid lines), the hatched areas indicate the projection of the bulk bands. The surface-state bands appear in the band gap and intersect each other at the Dirac point, forming a Dirac cone (shaded region, marking the analytical expression).} 
    \label{fig:spec}
\end{figure*}

Our aim is to obtain a model that accurately describes the topological surface states and can be discretized on a lattice using the finite-difference method~\cite{datta_2005}, without the appearance of unphysical states in the bulk band gap. As the distinctive feature of 3D TIs is the Dirac cone of the topological surface states, the region in $k$-space around $\Gamma$ where this Dirac cone appears is of central importance.
Thus, the material parameters used in Eq.~\eqref{eq:H_3D} should yield a dispersion that reliably describes the bulk bands, up to the momenta where the Dirac cone merges with the bulk bands. The first full set of parameters introduced for the Bi$_2$Se$_3$ family of materials in Ref.~\cite{Liu2010B} were obtained using perturbation theory, giving an accurate description of the low energy states only very close to $\Gamma$ 
(up to $k_{x,z}<0.04~\text{\AA}^{-1}$).
In a later work, a $\boldsymbol{\mathrm{k\cdot p}}$ perturbation approach applied to \emph{ab initio} calculations yielded a good qualitative description of the conduction (CB) and valence bands (VB) of Bi$_2$Se$_3$ and Sb$_2$Te$_3$~\cite{nechaev2016}. However, for Bi$_2$Te$_3$, in the region in $k$-space where the Dirac cone appears, the obtained dispersion has a significantly smaller band gap than the one observed in \emph{ab initio} band structures (see Figs.~\ref{fig:spec}d,e). In our fitting procedure the eigenvalues of the bulk model Hamiltonian~\eqref{eq:H_3D} are fitted to \emph{ab initio} band structures of bulk Bi$_2$Se$_3$, Bi$_2$Te$_3$ and Sb$_2$Te$_3$, with the band structure of Bi$_2$Te$_3$ shown in Fig.~\ref{fig:spec}b. More details on the \emph{ab initio} calculations are given in Appendix \ref{app:ab_init}. To obtain the most accurate fit, we vary the region in $k$-space over which the CB and VB are considered (imposing a minimal extent of the region to accurately capture the Dirac cone in the bulk gap), and we maximize the coefficient of determination $R^2$ of the fit. The conditions $|C_1|<|M_1|$, $|C_2|<|M_2|$, $M_0 M_1<0$ and $M_0M_2<0$ (see Sec.~\ref{sec:models}A for more details) were enforced on the fitted parameters of the bulk model. The resulting parameters are shown in Table~\ref{tab:param_fit} alongside the parameters of Refs.~\cite{Zhang2009,Liu2010B,nechaev2016}. For more details on the fitting procedure, see Appendix~\ref{app:fit}. We also evaluate the size of the band gap as the difference between the minimum of the CB and the maximum of the VB. In Figs.~\ref{fig:spec}d,e, we show the result of our fit for Bi$_2$Te$_3$ to the relevant bands, together with the dispersion obtained with parameters taken from Refs.~\cite{Liu2010B, nechaev2016}. The band gap of the fitted dispersion is much closer to that of the \emph{ab initio} calculation, than the value obtained through perturbation theory.

\begin{figure*}
    \centering
    \includegraphics[width=\linewidth]{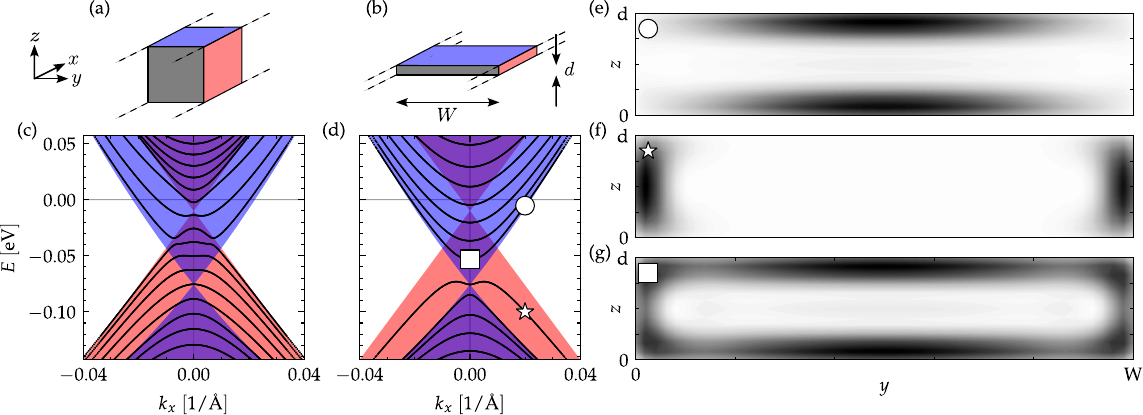} 
    \caption{(a),(b) Schematics of the nanostructures under consideration: (a) a nanowire with $W \sim d$ and (b) a nanoribbon with $W\gg d$.
    (c),(d) The spectrum of (c) a nanowire with $W=d=28$ nm and (d) a nanoribbon with $W=50$ nm and $d=6$ nm, with the fitted material parameters for Sb$_2$Te$_3$. The Dirac cone of the surface state in the $x\text{-}y$ ($x\text{-}z$) plane (see Figs.~\ref{fig:spec}f,g) is shown with blue (red) shading.
    (e),(f),(g) The wave function density $|\Psi(y,z)|^2$ over the nanoribbon cross section of states localized in different Dirac cones [see corresponding symbol in (d)], indicated by the grayscale colormap (arbitrary units). Where both Dirac cones overlap (purple shading), the states are delocalized over the whole perimeter of the nanoribbon as shown in (g).}
    \label{fig:slab_nanor} 
\end{figure*} 

\subsection{Surface-state model}

When confined to a semi-infinite geometry, with a surface in the $x\text{-}y$ plane at $z=0$, the bulk model Hamiltonian \eqref{eq:H_3D} yields a gapless surface-state spectrum, described by the following effective Hamiltonian~\cite{shen_book, Shan_2010, brey2014}, 
\begin{equation}
\begin{split}
   &H_\text{surf}^{\hat{z}}(k_x, k_y) = C_0 - C_1 M_0 /M_1 \label{eq:H_surf} \\ &\quad - \text{sgn}(M_1 )\sqrt{ 1 - (C_1 /M_1 )^2} A_0 (k_x\sigma_y - k_y\sigma_x ).
\end{split}
\end{equation}
The wave function profile perpendicular to the $x\text{-}y$ surface of the $k_x=k_y=0$ surface state has the following form when the 3D TI is confined to the $z>0$ region~\cite{shen_book,Liu2010B}
\begin{eqnarray}
    \chi(z) &=&\begin{pmatrix}
        c_1&-c_1&c_2&c_2
    \end{pmatrix}^T(e^{q_z^+z} - e^{q_z^-z}),\label{eq:wf_surf1} \\
    q_z^\pm &\equiv& \frac{1}{2}\sqrt{\dfrac{B_0^2}{M_1^2-C_1^2}}\pm\sqrt{\dfrac{1}{4}\dfrac{B_0^2}{M_1^2-C_1^2}+\dfrac{M_0}{M_1}},
   \label{eq:wf_surf2}        
\end{eqnarray}
with two independent parameters $c_1$ and $c_2$ (up to normalization). The wave function extends into the bulk with a characteristic penetration depth  $\lambda_z=\text{max}\{1/\mathfrak{R}(q_z^+),1/\mathfrak{R}(q_z^-)\}$~\cite{moors2018}, which is listed in Table~\ref{tab:param_fit} for the different sets of material parameters. In Fig.~\ref{fig:spec}f the solutions of the Hamiltonian~\eqref{eq:H_surf} (dashed lines) are shown, along with the numerically evaluated spectrum of a semi-infinite slab (black lines), which we obtain by using a version of the bulk Hamiltonian~\eqref{eq:H_3D} discretized on a lattice with confinement along $z$ and translational invariance along $x$ and $y$. This solution is for a surface orthogonal to the $\hat{z}$ direction, and analogous solutions can be obtained for surfaces with other orientations. In Fig.~\ref{fig:spec}g the solutions for a surface over the $x\text{-}z$ plane at $y=0$ is given (dashed lines), together with the numerical result (black lines). For simplicity, we set $C_0=0$ in all calculations, since this term only yields an overall shift of the spectrum in energy.

Note that in general, while barely noticeable in Figs.~\ref{fig:spec}f,g, there can be a shift in energy between the Dirac points (DPs) of the surface states, i.e., the energy where the surface-state bands cross each other at $\Gamma$. In the Dirac-like dispersion given by Eq.~\eqref{eq:H_surf}, the DP can be defined as $E_\text{DP}^{\hat{z}}=C_0-M_0 C_1/M_1$. Surfaces orthogonal to the $\hat{x}$ or $\hat{y}$ direction will host surface states with the DP positioned at $E_\text{DP}^{\hat{x}/\hat{y}}=C_0-M_0 C_2/M_2$. We define the difference between the two DPs as $\Delta E_\text{DP} = |E_\text{DP}^{\hat{x}/\hat{y}}-E_\text{DP}^{\hat{z}}|$. For the studied materials (Bi$_2$Se$_3$, Bi$_2$Te$_3$, Sb$_2$Te$_3$), we obtain $\Delta E_\text{DP}=24, 12$ and $ 56~\text{meV}$ from the fitted parameters, respectively, as also shown in Table~\ref{tab:param_fit}. When the electron-hole asymmetry is ignored ($C_1=C_2=0$), or the parameters are considered to be isotropic ($C_1=C_2$ and $M_1=M_2$) one naturally obtains $\Delta E_\text{DP} = 0$ ($E_\text{DP}^{\hat{z}}=E_\text{DP}^{\hat{x}/\hat{y}}$). 
However, as the material parameters in Table~\ref{tab:param_fit} show, anisotropy and electron-hole asymmetry can be significant.

Another important consequence of anisotropy and electron-hole asymmetry is that the DP is not centered in the middle of the bulk gap.
This can also be seen in \emph{ab initio} calculations of slab geometries or angle-resolved photoemission spectroscopy measurements~\cite{Zhang2010A, Liu2012, Jiang2012}. In Table~\ref{tab:param_fit} we evaluate the relative position of the DP for a surface in the $x\text{-}y$ plane as 
\begin{equation}
    \zeta_{\text{DP}} = \frac{E_\text{DP}^{\hat{z}}-E_{\text{VB}}(\Gamma)}{E_{\text{CB}}(\Gamma)-E_{\text{VB}}(\Gamma)}=\frac{1}{2}\bigg(1-\text{sign}(M_0)\frac{C_1}{M_1}\bigg),
    \label{eq:zeta_DP}
\end{equation}
where $E_{\text{VB(CB)}}(\Gamma)$ is the energy of the VB (CB) at $\Gamma$. A DP centered between the VB and CB at $\Gamma$ yields $\zeta_\text{DP} = 0.5$, while a DP at the top (bottom) of the VB (CB) at $\Gamma$ corresponds to $\zeta_{\text{DP}}=0~(1)$.

\subsection{Thin-film model}

In the thin-film limit, with surfaces in the $x\text{-}y$ plane at $z=0$ and $z=d$ (see Fig.~\ref{fig:slab_nanor}b), tunneling between the surface states on the top and bottom surfaces can open a finite \emph{hybridization gap} in the Dirac cone of the surface states. In this limit, the dispersion of the system can be captured using a low-energy thin-film model Hamiltonian~\cite{Lu2010, Shan_2010, shen_book},
\begin{eqnarray}
    H_\text{tf}(k_x,k_y) &=& E_0-Dk_{\parallel}^2+\hbar v_\text{F}(k_y\sigma_x-k_x\sigma_y)\nonumber\\
    &&+\,(\Delta/2-Bk_{\parallel}^2)\sigma_z\tau_z,
    \label{eq:H_2D}
\end{eqnarray}
with $\sigma_i$ ($i=x,y,z$) still the Pauli matrices for spin, $k_{\parallel}^2 \equiv k_x^2+k_y^2$, and $\tau_z$ acting on a different subspace from the one before in Eq.~\eqref{eq:H_3D}, with eigenvalues $\pm$ representing a hyperbola index that distinguishes between the doubly-degenerate surface-state solutions of Eq.~\eqref{eq:H_2D}. The Hamiltonian of Eq.~\eqref{eq:H_2D} is equivalent (ignoring the term $-D k_{\parallel}^2$) to the four-band effective model for a two-dimensional QSHI proposed by Bernevig, Hughes and Zhang~\cite{bhz}, which has been shown to capture the behavior of the bulk model of Eq.~\eqref{eq:H_3D} in the thin-film limit~\cite{Liu2010A}. The hybridization gap of the surfaces states is $\Delta E_\text{hyb}=2\Delta$ in this model, while the gap is trivial if $\Delta B<0$, and the system is in the nontrivial inverted regime if $\Delta B>0$, yielding a QSHI state~~\cite{bhz, Lu2010}. A non-zero value of $D$ results in electron-hole asymmetric surface states and $v_\text{F}$ is the Fermi velocity of the surface states. It should be noted that $|D|<|B|$ is required, otherwise there is no band gap at large $|\mathbf{k}|$~\cite{Lu2010, shen_book}.

\section{Nanostructures}
\label{sec:nano}

In this section, we consider the bulk model Hamiltonian~\eqref{eq:H_3D} to study the dispersion of two relevant 3D TI nanostructures: nanowires, with approximately equal width and thickness ($W\sim d$), and nanoribbons, for which $W\gg d$, shown schematically in Figs.~\ref{fig:slab_nanor}a,b.
We use Kwant~\cite{Groth2014} to obtain the spectra of the nanostructures, and Adaptive~\cite{Nijholt2019} for efficient parameter sampling. Note that we describe thicknesses in terms of QL and nm interchangeably, as 1 QL $\approx$ 1 nm for the materials under consideration.

Experimentally, 3D TI films down to a few-QL thickness have been achieved in all three materials~\cite{Zhang2010A, Zhang2011, Liu2012,Jiang2012, Vidal_2013, Landolt_2014, Neupane_2014} and nanoribbons with widths down to 50 nm were realized~\cite{kong2010, Xiu_2011}. Here we consider nanostructures with negligible hybridization between surface states on opposing sides (for the thin-film limit where hybridization becomes relevant, see Sec.~\ref{sec:thin-film}). In Fig.~\ref{fig:slab_nanor}c, we present the dispersion of a nanowire with square cross section and $W=d=28$ nm long edges (black lines). In Fig.~\ref{fig:slab_nanor}d, we also present the dispersion of a nanoribbon with the same perimeter as the nanowire, but a much larger width-to-height ratio, $W=50$ nm and $d=6$ nm. Here we have chosen Sb$_2$Te$_3$ because it has the largest value of $\Delta E_{\text{DP}}$ for the fitted parameters. The obtained spectra qualitatively resemble a conventional Dirac cone with confinement quantization~\cite{Bardarson2010, Zhang2010B}. However, there is a clear difference between the dispersions of the nanowire and the nanoribbon. The differences can be attributed to the effect of the surfaces of the nanostructures that are oriented in different directions, which we explain below.

To understand the effect of the different sides, in Figs.~\ref{fig:slab_nanor}c,d, we overlay the Dirac cones of the top/bottom (left/right side) surfaces, centered around their respective DPs, with a blue (red) shading. The states appearing in the spectrum of the nanostructures can be divided into three groups: states with energies in the blue regions, extending over the top and bottom surfaces (Fig.~\ref{fig:slab_nanor}e); states with energies in the red regions that are localized on the side surfaces (Fig.~\ref{fig:slab_nanor}f); and states that wrap around the whole perimeter of the nanowire in the regions where both Dirac cones overlap (Fig.~\ref{fig:slab_nanor}g). As the side surfaces of the nanoribbon have a much smaller area than the top and bottom surfaces, the side surfaces do not host as many states as in the case of the nanowire. Hence, the spectrum of the nanoribbon resembles more closely a quantized Dirac cone as expected for the top and bottom surface.

A relevant energy scale in the nanostructure is the spacing of subbands originating from the confinement of the surface states to the finite perimeter $P$ of its cross section. In the case of a Dirac dispersion, the spacing of the subbands can be approximated by $2\pi v_\text{F}/P$. For the nanoribbon considered in Fig.~\ref{fig:slab_nanor}d, one obtains $2\pi v_\text{F}/P=17~\text{meV}$ (here we consider an isotropic Fermi velocity $\hbar v_\text{F}\approx3~\text{eV\AA}$). The spacing of the subbands has to be compared to the effect of the surfaces of the nanoribbon being oriented in different directions, which can be quantified using the energy difference of DPs $\Delta E_{\text{DP}}$, which is 56 meV in the case of Sb$_2$Te$_3$. From this observation, we can deduce that the difference in DP energy on different sides may even affect the surface-state spectrum of the smallest attainable nanostructures.

\section{Thin-film limit}
\label{sec:thin-film}

We first investigate the thin-film limit in Sec.~\ref{subsec:treatment-bulk}, using a discretized version of the bulk model Hamiltonian~\eqref{eq:H_3D} and we compare the results to experimental findings, with an emphasis on the topology and size of the hybridization gap. Second, in Sec.~\ref{subsec:treatment-thin-film}, we consider a different approach to capture the physics of the thin-film limit quantitatively. Material parameters for the thin-film model Hamiltonian~\eqref{eq:H_2D} are obtained by fitting the model to band structure data of \emph{GW} calculations that accurately describe the thin-film limit~\cite{Forster2015,Forster2016,Forster_phd}.

\subsection{Treatment with the bulk model}
\label{subsec:treatment-bulk}

\begin{figure}[t]
    \centering
    \includegraphics[width=\linewidth]{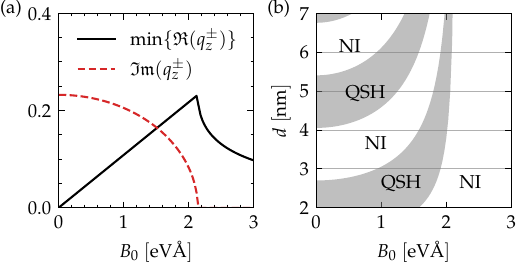}        
    \caption{(a) The real and imaginary part of $q_z^\pm$ (see Eq. \eqref{eq:wf_surf2}) as a function of the out-of-plane Fermi velocity $B_0$. The real part is inversely proportional to the penetration depth of the surface state, while the imaginary part of $q_z^\pm$ determines the topological phase at a given thickness $d$. (b) When described using the bulk model~\eqref{eq:H_3D} the topological phase of the 3D TI slab oscillates between a NI and QSHI as the thickness $d$ is varied. The period of this oscillation increases as a function of $B_0$ and diverges above a critical value of $B_0$ such that the slab is always in the NI state.}
    \label{fig:oscill}
\end{figure}

When described using the bulk model of Eq.~\eqref{eq:H_3D}, it was shown that the hybridization gap in the thin-film limit oscillates in size as the thickness is varied. The closing and subsequent reopening of hybridization gap occurs at certain critical thicknesses $d_{\text{c}n}$ (see Fig.~\ref{fig:oscill}b). The low-energy physics of the system at a thickness close to $d_{\text{c}n}$ can be described by the BHZ model~\cite{Liu2010A,bhz}, which implies that the oscillation of the hybridization gap is also accompanied by topological phase transitions, with the system alternating between a NI and a QSHI phase.

In a 3D TI thin film of thickness $d$ the interference of two transverse wave functions located on the top and bottom surfaces (orthogonal to $\hat{z}$), given by Eqs.~\eqref{eq:wf_surf1} and~\eqref{eq:wf_surf2}, will close the hybridization gap if the thickness matches the critical value $d_{\text{c}n}=n l_\text{c}$, where $l_\text{c} \simeq \pi/\mathfrak{Im}(q_z^\pm)$ is the period of the oscillation~\cite{Shan_2010}.
When the out-of-plane Fermi velocity $B_0$ is set to zero, the period of the oscillation is equal to $l_\text{c}(B_0=0)=\pi\sqrt{|M_0/M_1|}$. As shown in Fig.~\ref{fig:oscill} for Bi$_2$Te$_3$, when $B_0$ takes on a non-zero value, the value of $l_\text{c}$ increases from $l_\text{c}(B_0=0)$. However, if $B_0\gtrsim 2\sqrt{-M_0(M_1^2-C_1^2)/M_1}$, the imaginary part of $q_z^\pm$ goes to zero, the oscillatory behavior vanishes and the system is a NI for all thicknesses.

\begin{figure*}             
    \includegraphics[width=\linewidth]{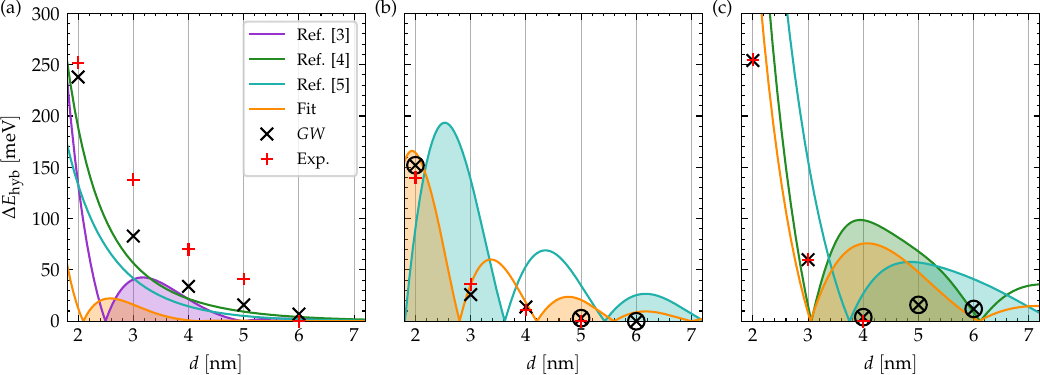}
    \caption{Size of the hybridization gap $\Delta E_{\text{hyb}}$ at the $\Gamma$ point for thicknesses $2$ nm $<d<7$ nm for (a) Bi$_2$Se$_3$, (b) Bi$_2$Te$_3$ and (c) Sb$_2$Te$_3$. The gap is extracted from the spectra of the surface states in a slab geometry described using the bulk model~\eqref{eq:H_3D}, with parameters taken from Refs.~\cite{Zhang2009, Liu2010B, nechaev2016} and our fits. At a given thickness the system is a QSHI (NI) if the area below the curve is colored (blank). The hybridization gap obtained in the $GW$ calculations~\cite{Forster2015, Forster2016} are included, with the QSHI states marked with a circle. For reference, experimentally determined gaps at the Dirac point are also shown~\cite{Zhang2010A, Liu2012, Jiang2012}.}
    \label{fig:hyb_gap}
\end{figure*}

In Fig.~\ref{fig:hyb_gap}, for Bi$_2$Se$_3$, Bi$_2$Te$_3$ and Sb$_2$Te$_3$, the numerically evaluated hybridization gap $\Delta E_\text{hyb}$ of the surface states at $\Gamma$ is given as a function of the film thickness, where we use a discretized version of the bulk model Hamiltonian~\eqref{eq:H_3D}. For the different parameters sets of Eq.~\eqref{eq:H_3D}, we also indicate the topological phase for every value of $d$ with blank (filled) shading below the curve for the NI (QSHI) phase. We compare the size of the hybridization gap of the bulk model Hamiltonian~\eqref{eq:H_3D} with values determined experimentally from angle-resolved photoemission spectroscopy measurements~\cite{Zhang2010A, Liu2012, Jiang2012}. For the topology of the gap we consult results obtained with the \emph{GW} method that accurately capture the many-body effects of the surface states ~\cite{Forster2015, Forster2016, Forster_phd}. The size of the hybridization gap as determined from the \emph{GW} calculations is also shown. The exact values of $\Delta E_{\text{hyb}}$ determined with the different methods can be found for 2-6 QL thicknesses in Table~\ref{tab:hyb_gap}, with a blank (grey) background indicating that the thin film is a NI (QSHI).

According to experimental measurements and \emph{GW} calculations, it is expected that Bi$_2$Se$_3$ is a NI in the thin-film limit~\cite{Forster2015,Liu2023,Wang2019}. The parameter set from Ref.~\cite{Zhang2009} and the fitted parameter set do not capture accurately this behavior, rather suggesting that the thin film is in the QSHI above a thickness of 3 QL. The parameter set listed in Refs.~\cite{Liu2010B,nechaev2016} gives a better description of Bi$_2$Se$_3$ thin films, yielding a NI for all thicknesses with hybridization gaps which are systematically smaller than the 
experimentally determined values, with Ref.~\cite{Liu2010B} giving the closest values. Remarkably, the hybridization gaps for Bi$_2$Te$_3$ match the experimentally determined values up to 4 QL very well when considering our parameter set obtained via fitting, whereas the topology of the gap is also captured up to 5 QL. In contrast, the same cannot be said of the parameters set of Ref.~\cite{nechaev2016}, while the parameter set of Ref.~\cite{Liu2010B} yields a gapless dispersion. In the case of Sb$_2$Te$_3$, all parameters sets capture the topology of the hybridization gap accurately, but the size of the hybridization gap deviates significantly from experimentally measured values. From the obtained hybridization gaps, the thin film is NI for 2 and 3 QL and QSHI for 4, 5 and 6 QL, as predicted by \emph{GW} calculations. While reasonable agreement is retrieved in certain cases, the results shown in Fig.~\ref{fig:hyb_gap} indicate that parameter sets, obtained with both perturbative and fitting approaches, cannot be used to describe the thin-film limit reliably for the different materials.

\subsection{Treatment with the thin-film model}
\label{subsec:treatment-thin-film}

As highlighted in the previous section, it is evident that the surface-state dispersion obtained when confining the bulk model~\eqref{eq:H_3D} to a slab geometry does not reliably capture the electronic and topological properties of thin films observed in experiment. However, recent theoretical works have shown that employing the \emph{GW} method in determining the  properties of thin films yields band structure data that is in excellent agreement with angle-resolved photoemission spectroscopy measurements~\cite{Forster2015,Forster2016}. By fitting the low-energy effective model~\eqref{eq:H_2D} to this band structure data, we can obtain material parameters for the Bi$_2$Se$_3$ family of materials in the thin-film limit. Note that, by considering model Hamiltonian~\eqref{eq:H_2D}, the physics of the side surfaces, as discussed in the previous section (e.g., the difference in DP energy) is no longer captured.

\begin{figure}[t]
    \centering
    \includegraphics[width=\linewidth]{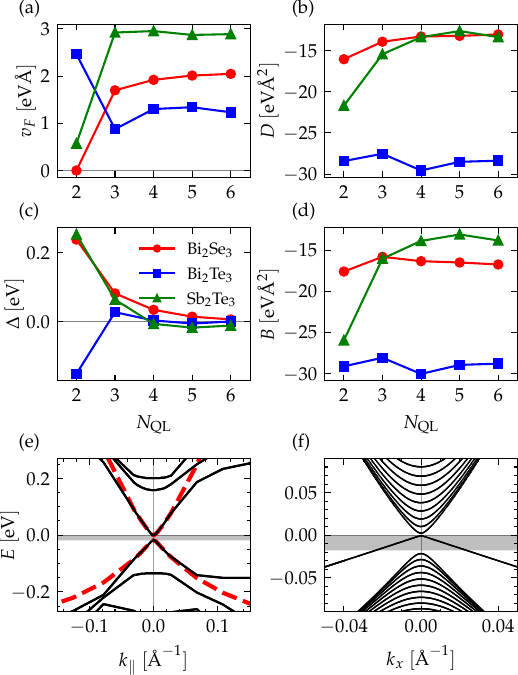}
    \caption{(a),(b),(c),(d) The parameters of the effective thin-film model \eqref{eq:H_2D} for film thicknesses from 2 to 6 QLs determined by fitting to spectra obtained with the \emph{GW} method~\cite{Forster2015, Forster2016}. The parameters are obtained by imposing $|D|<|B|$, and $\Delta\cdot B <(>)\, 0$ when the thin film is a NI (QSHI). (e) Band structure of a 5 QL thick Sb$_2$Te$_3$ thin-film obtained with the \emph{GW} method (continuous black line), and the fitted dispersion (red dashed line). The shaded region corresponds to the hybridization gap $\Delta E_{\text{hyb}}$. (f) The spectrum of a Sb$_2$Te$_3$ nanoribbon with $W=100$ nm and $d=5$ QL. The parameters were taken from Table~\ref{tab:tf_params}.}
    \label{fig:eff_fit}
\end{figure}

The material parameters were obtained for thicknesses varying between 2 and 6 QL. We imposed the constraint $|D|<|B|$ such that the obtained parameters yield a gapped dispersion. It has been shown that one can obtain parameters corresponding to a QSHI or a NI from fitting to the same surface-state spectrum without any discernible difference~\cite{Forster2015}. Thus, we take the liberty of imposing the constraint $\Delta B > 0$ when the $GW$ results suggest that the system is a QSHI and conversely impose $\Delta B < 0$ when the system is a NI~\cite{Forster2016}. The obtained parameters are shown in Figs.~\ref{fig:eff_fit}a,b,c,d and listed in Table~\ref{tab:tf_params}. In Fig.~\ref{fig:eff_fit}e the \emph{GW} band structure of 5 QL thick Sb$_2$Te$_3$ thin-film is presented, together with the fitted dispersion, and in Fig.~\ref{fig:eff_fit}f the dispersion of a nanoribbon with the same thickness and width $W=100$ nm is shown. At this thickness, the Sb$_2$Te$_3$ thin-film is a QSHI, thus edge states appear in the dispersion of the nanoribbon with energies in the hybridization gap.

\section{Discussion}
\label{sec:disc}

Comparing the values in Tables~\ref{tab:param_fit} and \ref{tab:hyb_gap}, it is clear that there is substantial variation between all the parameter sets and their properties that are highly relevant for the low-energy spectrum (e.g., bulk gap, DP positioning and asymmetry). Hence, it can be of crucial importance to consider an appropriate parameter set that is tailored to the specific TI material for accurately modeling the nanoelectronic properties of TI nanostructure-based devices.

In this work, we focus on the accurate low-energy description of 3D TI nanostructures, including the thin-film limit with a hybridization gap, and the impact of anisotropy and electron-hole asymmetry. However, there are more aspects that are not under consideration in this work while being relevant for nanoelectronic device modeling.
Aside from a thickness-dependent hybridization gap, there is also a thickness-dependent energy shift of the spectrum, captured by thin-film model parameter $E_0$. For this shift, the $GW$ calculations that we considered for our model fits do not line up with experimental data, while other many-body calculations match experimental values better~\cite{Liu2023}. However, such a shift can be easily taken into account by adjusting the Fermi level in the nanostructure simulations accordingly.
Another important effect can be observed in 3D TI films grown by molecular beam epitaxy. The top surface is usually exposed to vacuum and the bottom surface is attached to a substrate, breaking the inversion symmetry along the $\hat{z}$ direction. In this case, a structure inversion asymmetry term can account for the influence of the substrate~\cite{Shan_2010, Lu_2013}. Such effects can also be revealed through asymmetric electrostatic gating~\cite{ziegler_2018, rosenbach_2022}. While not under consideration here, we can also introduce such terms in a straightforward manner in our models.

\begin{table}[t]
    \centering
    \begin{tabular}{ >{\centering\arraybackslash}m{5ex} >{\arraybackslash}m{22ex} | >{\centering\arraybackslash}m{4ex}>{\centering\arraybackslash}m{4ex}>{\centering\arraybackslash}m{4ex}>{\centering\arraybackslash}m{4ex}>{\centering\arraybackslash}m{4ex} }
    \toprule
    &&\multicolumn{5}{c}{$N_{\text{QL}}$}\Tstrut\\
    &&2&3&4&5&6\\
    \hline
    \parbox[t]{8mm}{\multirow{11}{*}{\rotatebox[origin=c]{90}{$\Delta E_{\text{hyb}}~[\text{meV}]$}}}
    &Bi$_2$Se$_3$~\cite{Zhang2009}&137 & \cellcolor{gray!50}40 & \cellcolor{gray!50}22  & - & - \Tstrut\Bstrut\\    
    &Bi$_2$Se$_3$~\cite{Liu2010B} &191 & 58 & 22 & 10 & 4 \\
    &Bi$_2$Se$_3$~\cite{nechaev2016} &133 & 42 & 15 & 5 & - \\
    &Bi$_2$Se$_3$ fit            &13 & \cellcolor{gray!50}16 & - & - & - \\
    \hhline{*{2}{|~}*{5}{|-}}
    &Bi$_2$Se$_3$ \emph{GW}~\cite{Forster2015} &238&83&34&(16)&(7) \Tstrut\\
    \hhline{*{1}{|~}*{6}{|-}} 
    &Bi$_2$Te$_3$~\cite{Liu2010B} &-&-&-&-&-\Tstrut\\
    &Bi$_2$Te$_3$~\cite{nechaev2016}& \cellcolor{gray!50}91 & \cellcolor{gray!50}138 & 54 & 35 & \cellcolor{gray!50}25 \\
    &Bi$_2$Te$_3$ fit            & \cellcolor{gray!50}161 & 50& 6 & \cellcolor{gray!50}18 & 11\\
    \hhline{*{2}{|~}*{5}{|-}}
    &Bi$_2$Te$_3$ \emph{GW}~\cite{Forster2016,Forster_phd}& \cellcolor{gray!50} 152&26&(14)& \cellcolor{gray!50} (3)& \cellcolor{gray!50} (0)\Tstrut\\
    \hhline{*{1}{|~}*{6}{|-}}
    &Sb$_2$Te$_3$~\cite{Liu2010B} &536 & 15 & \cellcolor{gray!50} 99 & \cellcolor{gray!50} 68 & \cellcolor{gray!50} 9  \Tstrut\\ 
    &Sb$_2$Te$_3$~\cite{nechaev2016} & 747 & 156 & \cellcolor{gray!50} 27 &\cellcolor{gray!50} 57&\textbf{\cellcolor{gray!50}} 41 \\
    &Sb$_2$Te$_3$ fit            &587 & 78 & \cellcolor{gray!50} 65 & \cellcolor{gray!50} 73 &  \textbf{\cellcolor{gray!50}} 30 \\
    \hhline{*{2}{|~}*{5}{|-}}
    &Sb$_2$Te$_3$ \emph{GW}~\cite{Forster2016,Forster_phd}&254&60& \cellcolor{gray!50} (4)& \cellcolor{gray!50} (16)& \cellcolor{gray!50} (12)\Tstrut\\
    \toprule 
    \end{tabular}
    \caption{The size of the hybridization gap $\Delta E_\text{hyb}$ of Bi$_2$Se$_3$, Bi$_2$Te$_3$ and Sb$_2$Te$_3$ thin films, described using the bulk model Hamiltonian~\eqref{eq:H_3D} for different sets of material parameters and varying number of quintuple layers $N_{\text{QL}}$. For each case, the topological phase is indicated by a gray (blank) shading if the thin film is a QSHI (NI). The results of the \emph{GW} calculations are also given. The gaps in parentheses should not be taken at face value, as the accuracy of the \emph{GW} calculations does not allow a safe statement when the band gap becomes too small~\cite{Forster_phd}. }
    \label{tab:hyb_gap} 
\end{table}

In addition to offering a good starting point for nanoelectronic device modeling, the models considered in this work are also suitable for the study of hybrid devices that include superconductivity, since an extension to a Bogoliubov-de Gennes framework follows naturally~\cite{sitthison2014, Heffels2023}. The effect of proximity-induced superconducting pairing in 3D TIs has received considerable interest~\cite{Fu2008, Stanescu2010, potter2011, cook2011, cook2012, sitthison2014,chiu2016,Bai_2022}, with the effective thin-film model~\eqref{eq:H_2D} being considered for the study of (proximitized) magnetically doped 3D TI nanoribbons~\cite{Chen2018,Zeng2018,Atanov2024, Burke2024}. In such systems, typically electron-hole asymmetry is neglected ($D=0$). It has been shown, however, that electron-hole asymmetry can play an important role in the transport properties of magnetically doped 3D TI films~\cite{Lu_2013}.

Here, we focus on obtaining suitable model parameters for Bi$_2$Se$_3$, Bi$_2$Te$_3$ and Sb$_2$Te$_3$ by considering four-band models and a parameter fitting strategy instead of perturbation theory applied at $\Gamma$. Alternatively, however, it has already been shown that an eight-band model Hamiltonian obtained from perturbation theory can also give an accurate low-energy description of all three materials~\cite{nechaev2016}. However, as that approach is also expected to become unreliable in the thin-film limit and is computationally more demanding, our approach offers some distinct advantages for efficient and accurate nanoelectronic device modeling.

\section{Conclusion}
\label{sec:conc}

We model the low-energy electronic spectrum of 3D TI nanostructures (e.g., nanowires and nanoribbons) based on Bi$_2$Se$_3$, Bi$_2$Te$_3$ and Sb$_2$Te$_3$ in detail. We use the commonly considered four-band (bulk and thin-film) model Hamiltonians and derive new parameter sets by fitting to \emph{ab initio} band structure data. 
Our fitting strategy is tailored to accurately capture the (in general, anisotropic and electron-hole asymmetric) low-energy electronic structure of the Dirac surface states, while avoiding any unphysical behavior that may arise when the Hamiltonian is discretized on a lattice. We studied the accuracy of the obtained fitted material parameters in the thin-film limit, when the surface states hybridize through the bulk, by using a discretized version of the bulk Hamiltonian. We have found that our fitting method yields material parameters that capture the size and topology of the hybridization gap in Bi$_2$Te$_3$ remarkably well. However, both our new and existing sets of parameters cannot reliably describe the thin-film limit for all thicknesses and materials under consideration. Hence, we resort to a thin-film model with material parameters extracted from the surface-state spectra of thin-film \emph{GW} calculations. With our new parameter sets, the considered models provide a suitable framework for simulating the low-energy spectrum and corresponding properties (e.g., topology, transport) of 3D TI-based nanoelectronic devices with a broad range of applications.

\begin{acknowledgments}
We thank Peter Krüger for providing the band structure data of the \emph{GW} calculations in the thin-film limit. We also thank Jonas Buchhorn, Malcolm Connolly, Stefanos Dimitriadis, Dennis Heffels, Jan Karthein, Jalil Abdur Rehman, Thomas Schäpers, Michael Schleenvoigt, Peter Schüffelgen, Kaycee Underwood and Max Vaßen-Carl for useful discussions. This work is supported by the QuantERA grant MAGMA (by the German Research Foundation under grant 491798118, by the National
Research Fund Luxembourg under Grant No. INTER/QUANTERA21/16447820/MAGMA, and by MCIN/AEI/10.13039/501100011033 and the European Union NextGenerationEU/PRTR under project PCI2022-132927), and by Germany's Excellence Strategy -- Cluster of Excellence Matter and Light for Quantum Computing (ML4Q) EXC 2004/1 -- 390534769. K.M.\ and P.R.\ acknowledge the financial support by the Bavarian Ministry of Economic Affairs, Regional Development and Energy within Bavaria’s High-Tech Agenda Project ``Bausteine für das Quantencomputing auf Basis topologischer Materialien mit experimentellen und theoretischen Ansätzen'' (Grant No.\ 07 02/686 58/1/21 1/22 2/23), and K.M.\ acknowledges the financial support by the Quantum Future project ‘MajoranaChips’ (Grant No. 13N15264) within the funding program Photonic Research Germany.
\end{acknowledgments}

\appendix

\section{\emph{ab initio} calculations}
\label{app:ab_init}

In our density functional theory (DFT) calculations we use the full-potential relativistic Korringa-Kohn-Rostoker Green function method (KKR)~\cite{Ebert2011} as implemented in the JuKKR code~\cite{jukkr2022}. Our calculations are carried out for the experimental crystal structures for Bi$_2$Te$_3$~\cite{Bi2Te3}, Sb$_2$Te$_3$~\cite{Sb2Te3} and Bi$_2$Se$_3$~\cite{Bi2Se3} and we parametrize the exchange correlation functional using the local density approximation (LDA)~\cite{Vosko1980} because a comparison of LDA and generalized gradient approximation (using the PBE functional~\cite{PBE}) resulted in gap sizes that reproduce the experimentally observed gap better. We employ Lloyd's formula~\cite{Zeller2004} to correct for the error arising from the finite $\ell_\text{max}=3$ cutoff in the angular momentum expansion of the space filling Voronoi cells around the atomic centers, where the exact (i.e.\ full-potential) description of the atomic shapes is taken into account~\cite{Stefanou1990,Stefanou1991}. The DFT calculations are orchestrated using the AiiDA-KKR plugin~\cite{aiida-kkr-paper} to the AiiDA infrastructure for automated FAIR data provenance tracking~\cite{aiida}. Our results are uploaded to the Materials cloud archive~\cite{dataset} and the JuKKR and AiiDA-KKR codes are published as open-source software~\cite{jukkr2022, aiida-kkr-code}. 

\section{Fitting procedure}
\label{app:fit}

The material parameters for the bulk model (Eq.~\eqref{eq:H_3D}) and the effective thin-film model (Eq.~\eqref{eq:H_2D}) for the three studied materials are obtained by fitting to \emph{ab initio} band structures.

\begin{figure}[t]
    \centering
    \includegraphics[width=0.63\linewidth,trim={0 .9cm 0 0},clip]{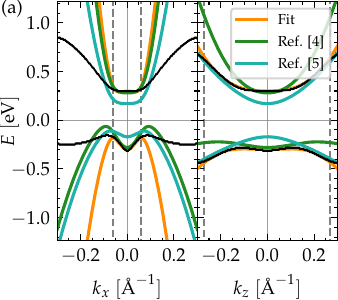}
    \includegraphics[width=0.35\linewidth,trim={0 1cm 0 0},clip]{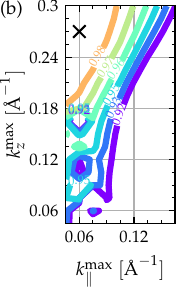}
    \includegraphics[width=0.63\linewidth]{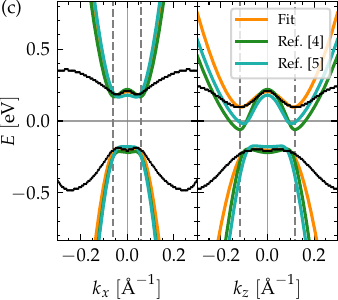}
    \includegraphics[width=0.35\linewidth]{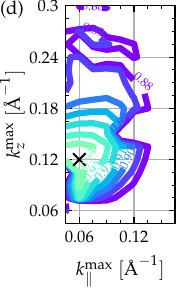}
    \caption{(a),(c) The dispersion resulting from the fit (orange) to \emph{ab initio} band structures (black dots), for (a) Bi$_2$Se$_3$ and (c) Sb$_2$Te$_3$, compared with dispersion evaluated with the material parameters taken from Ref.~\cite{Liu2010B} (green) and Ref.~\cite{nechaev2016} (cyan).  The upper limits of the fit in momentum space $k_{\parallel,z}^{\text{max}}$ are shown with gray dashed lines. (b),(d) A map of the $R^2$ of the fits for (b) Bi$_2$Se$_3$ and (d) Sb$_2$Te$_3$, obtained for different values of $k_\parallel^{\text{max}}$ and $k_{z}^{\text{max}}$. A black cross marks the selected pair of ($k_\parallel^{\text{max}}$, $k_{z}^{\text{max}}$) values.}
    \label{fig:fit_comparison}    
\end{figure}

\begin{table}[t]    
    \centering
    \begin{tabular}{c|c|ccccc}
    \toprule
     &  $N_{\text{QL}}$ & $E_0~[\text{eV}]$ & $D~[\text{eV\AA}^2]$ & $\Delta~[\text{eV}]$ & $B~[\text{eV\AA}^2]$ & $v_\text{F}~[\text{eV\AA}]$ \\
    \hline
    \multirow{5}{*}{Bi$_2$Se$_3$}
    &     2 &  0.121 & -16.14 &  0.239 & -17.56 & -0.048 \\
    &     3 &  0.043 & -13.94 &  0.082 & -15.82 &  1.697 \\
    &     4 &  0.018 & -13.31 &  0.034 & -16.35 &  1.920 \\
    &     5 &  0.008 & -13.23 &  0.014 & -16.51 &  2.010 \\
    &     6 & -0.002 & -13.06 &  0.006 & -16.75 &  2.046 \\
     \hline
    \multirow{5}{*}{Bi$_2$Te$_3$}
    &     2 &  0.077 & -28.41 & -0.153 & -29.14 &  2.463 \\
    &     3 &  0.013 & -27.52 &  0.027 & -28.10 &  0.876 \\
    &     4 &  0.001 & -29.55 &  0.003 & -30.06 &  1.301 \\
    &     5 &  0.002 & -28.50 & -0.006 & -28.97 &  1.340 \\
    &     6 &  0.000 & -28.36 & -0.000 & -28.81 &  1.232 \\
     \hline
    \multirow{5}{*}{Sb$_2$Te$_3$}
    &     2 &  0.127 & -21.70 &  0.254 & -26.29 &  0.482 \\
    &     3 &  0.030 & -15.45 &  0.064 & -16.04 &  2.921 \\
    &     4 &  0.002 & -13.39 & -0.007 & -13.89 &  2.952 \\
    &     5 &  0.008 & -12.65 & -0.018 & -13.10 &  2.870 \\
    &     6 &  0.006 & -13.38 & -0.012 & -13.83 &  2.887 \\
    \toprule
    \end{tabular}
    \caption{Parameters of the thin-film Hamiltonian~\eqref{eq:H_2D}, for Bi$_2$Se$_3$, Bi$_2$Te$_3$ and Sb$_2$Te$_3$, for thicknesses ranging 2-6 QL.
}
    \label{tab:tf_params} 
\end{table}

In the case of the bulk model, the eigenvalues of the Hamiltonian were fitted to the relevant bands from $\Gamma\equiv0$ up to a certain $k_\parallel^{\text{max}}$ value in $k_x$ and $k_y$, and $k_{z}^{\text{max}}$ in $k_z$, with a sampling interval of $0.003~\text{\AA}^{-1}$ along the three axes (for reference, the distance to other symmetry points shown in Fig.~\ref{fig:spec}c is $0.882~\text{\AA}^{-1}$ for $\Gamma \textnormal{-} \text{L}$, $0.903~\text{\AA}^{-1}$ for $\Gamma \textnormal{-} \text{F}$, and $0.329~\text{\AA}^{-1}$ for $\Gamma \textnormal{-} \text{Z}$). For each fit corresponding to a pair of $(k_\parallel^{\text{max}}, k_{z}^{\text{max}})$ values, the $R^2$ of the fit was evaluated (see Figs.~\ref{fig:fit_comparison}b,d). The best fit is obtained by choosing the one for which $R^2(k_\parallel^{\text{max}}, k_{z}^{\text{max}})>0.999\cdot\text{max}(R^2)$ and also $k_\parallel^{\text{max}}+k_{z}^{\text{max}}$ is maximal, while imposing certain constraints on $k_{z,\parallel}^\text{max}$, as explained below.
In Figs.~\ref{fig:fit_comparison}a,c, we show the result of our fits for Bi$_2$Se$_3$ and Sb$_2$Te$_3$ to the relevant bands, together with the dispersion obtained with parameters taken from Refs.~\cite{Liu2010B, nechaev2016}. In Figs.~\ref{fig:fit_comparison}b,d, maps of the $R^2$ values obtained for different pairs of $k_\parallel^{\text{max}}$ and $k_{z}^{\text{max}}$ are shown, with the optimal value indicated with a black cross. For the three materials the maximal value of $R^2$ was 0.9891, 0.9868 and 0.9458, for Bi$_2$Se$_3$, Bi$_2$Te$_3$ and Sb$_2$Te$_3$, respectively.

We impose a lower bound of $k_{z,\parallel}^{\text{max}} > 0.045~\text{\AA}^{-1}$ such that we capture at least the same extent as can be accurately captured with a perturbative approach~\cite{Zhang2010B}.
In order to have a fit that covers the whole region in $k$-space where the surface states can be found inside the bulk gap, we impose an additional constraint. The points in momentum space where the Dirac cone of the surface states joins the bulk-band energies can be approximated by
\begin{eqnarray}
    k_\parallel^\text{surf}=\frac{ \text{max}_{i \in \{0,1\}}\{E_i +(-1)^i (C_0- C_1M_0/M_1)\}}{A_0\sqrt{1-(C_1/M_1)^2}},\nonumber
\end{eqnarray}
for the in-plane dispersion, $E_0=-\text{max}_\mathbf{k}\{E_{\text{VB}}(\mathbf{k})\}$ and $E_1=\text{min}_\mathbf{k}\{E_{\text{CB}}(\mathbf{k})\}$. For completeness, the dispersion of the surface states with other orientations also have to be considered, e.g., orthogonal to the $\hat{y}$ direction. In this case, the Dirac cone will join the bulk-band energies at 
\begin{eqnarray}
    k_\parallel^\text{surf}=\frac{ \text{max}_{i \in \{0,1\}}\{E_i +(-1)^i (C_0- C_2M_0/M_2)\}}{A_0\sqrt{1-(C_2/M_2)^2}},\nonumber\\
    k_{z}^\text{surf}=\frac{ \text{max}_{i \in \{0,1\}}\{E_i +(-1)^i (C_0- C_2M_0/M_2)\}}{B_0\sqrt{1-(C_2/M_2)^2}},\nonumber
\end{eqnarray}
for the dispersion in $k_x$ and $k_z$, respectively. Using values found in literature, we found that $k_\parallel^\text{surf}, k_{z}^\text{surf}<0.2~\text{\AA}^{-1}$, which is contained in the region of $k$-space over which we sample the \emph{ab initio} band structures.
Hence, we can capture the surface states inside the bulk gap by imposing the conditions $k_\parallel^{\text{max}}>k_\parallel^\text{surf}$ and $k_{z}^{\text{max}}>k_{z}^\text{surf}$, where $k_{\parallel,z}^\text{surf}$ are evaluated using the \emph{ab initio} data and the fitted parameters.

For the effective surface-state model, a similar procedure was used, with the difference that the fit was performed for $k_x$ and $k_y$ up to $k_\parallel^{\text{max}}$, and the fit with $R^2>0.999\cdot\text{max}(R^2)$ and maximal $k_\parallel^{\text{max}}$ was selected.

\bibliographystyle{apsrev4-2}
\bibliography{main.bib}

\begin{thebibliography}{79}%
\makeatletter
\providecommand \@ifxundefined [1]{%
 \@ifx{#1\undefined}
}%
\providecommand \@ifnum [1]{%
 \ifnum #1\expandafter \@firstoftwo
 \else \expandafter \@secondoftwo
 \fi
}%
\providecommand \@ifx [1]{%
 \ifx #1\expandafter \@firstoftwo
 \else \expandafter \@secondoftwo
 \fi
}%
\providecommand \natexlab [1]{#1}%
\providecommand \enquote  [1]{``#1''}%
\providecommand \bibnamefont  [1]{#1}%
\providecommand \bibfnamefont [1]{#1}%
\providecommand \citenamefont [1]{#1}%
\providecommand \href@noop [0]{\@secondoftwo}%
\providecommand \href [0]{\begingroup \@sanitize@url \@href}%
\providecommand \@href[1]{\@@startlink{#1}\@@href}%
\providecommand \@@href[1]{\endgroup#1\@@endlink}%
\providecommand \@sanitize@url [0]{\catcode `\\12\catcode `\$12\catcode `\&12\catcode `\#12\catcode `\^12\catcode `\_12\catcode `\%12\relax}%
\providecommand \@@startlink[1]{}%
\providecommand \@@endlink[0]{}%
\providecommand \url  [0]{\begingroup\@sanitize@url \@url }%
\providecommand \@url [1]{\endgroup\@href {#1}{\urlprefix }}%
\providecommand \urlprefix  [0]{URL }%
\providecommand \Eprint [0]{\href }%
\providecommand \doibase [0]{https://doi.org/}%
\providecommand \selectlanguage [0]{\@gobble}%
\providecommand \bibinfo  [0]{\@secondoftwo}%
\providecommand \bibfield  [0]{\@secondoftwo}%
\providecommand \translation [1]{[#1]}%
\providecommand \BibitemOpen [0]{}%
\providecommand \bibitemStop [0]{}%
\providecommand \bibitemNoStop [0]{.\EOS\space}%
\providecommand \EOS [0]{\spacefactor3000\relax}%
\providecommand \BibitemShut  [1]{\csname bibitem#1\endcsname}%
\let\auto@bib@innerbib\@empty
\bibitem [{\citenamefont {Hasan}\ and\ \citenamefont {Moore}(2011)}]{hasan_2011}%
  \BibitemOpen
  \bibfield  {author} {\bibinfo {author} {\bibfnamefont {M.~Z.}\ \bibnamefont {Hasan}}\ and\ \bibinfo {author} {\bibfnamefont {J.~E.}\ \bibnamefont {Moore}},\ }\href {https://doi.org/10.1146/annurev-conmatphys-062910-140432} {\bibfield  {journal} {\bibinfo  {journal} {Annu. Rev. Condens. Matter Phys.}\ }\textbf {\bibinfo {volume} {2}},\ \bibinfo {pages} {55} (\bibinfo {year} {2011})}\BibitemShut {NoStop}%
\bibitem [{\citenamefont {Breunig}\ and\ \citenamefont {Ando}(2022)}]{Breunig2022}%
  \BibitemOpen
  \bibfield  {author} {\bibinfo {author} {\bibfnamefont {O.}~\bibnamefont {Breunig}}\ and\ \bibinfo {author} {\bibfnamefont {Y.}~\bibnamefont {Ando}},\ }\href {https://doi.org/10.1038/s42254-021-00402-6} {\bibfield  {journal} {\bibinfo  {journal} {Nat. Rev. Phys.}\ }\textbf {\bibinfo {volume} {4}},\ \bibinfo {pages} {184} (\bibinfo {year} {2022})}\BibitemShut {NoStop}%
\bibitem [{\citenamefont {Zhang}\ \emph {et~al.}(2009)\citenamefont {Zhang}, \citenamefont {Liu}, \citenamefont {Qi}, \citenamefont {Dai}, \citenamefont {Fang},\ and\ \citenamefont {Zhang}}]{Zhang2009}%
  \BibitemOpen
  \bibfield  {author} {\bibinfo {author} {\bibfnamefont {H.}~\bibnamefont {Zhang}}, \bibinfo {author} {\bibfnamefont {C.-X.}\ \bibnamefont {Liu}}, \bibinfo {author} {\bibfnamefont {X.-L.}\ \bibnamefont {Qi}}, \bibinfo {author} {\bibfnamefont {X.}~\bibnamefont {Dai}}, \bibinfo {author} {\bibfnamefont {Z.}~\bibnamefont {Fang}},\ and\ \bibinfo {author} {\bibfnamefont {S.-C.}\ \bibnamefont {Zhang}},\ }\href {https://doi.org/doi:10.1038/nphys1270} {\bibfield  {journal} {\bibinfo  {journal} {Nat. Phys.}\ }\textbf {\bibinfo {volume} {5}},\ \bibinfo {pages} {438 } (\bibinfo {year} {2009})}\BibitemShut {NoStop}%
\bibitem [{\citenamefont {Liu}\ \emph {et~al.}(2010{\natexlab{a}})\citenamefont {Liu}, \citenamefont {Qi}, \citenamefont {Zhang}, \citenamefont {Dai}, \citenamefont {Fang},\ and\ \citenamefont {Zhang}}]{Liu2010B}%
  \BibitemOpen
  \bibfield  {author} {\bibinfo {author} {\bibfnamefont {C.-X.}\ \bibnamefont {Liu}}, \bibinfo {author} {\bibfnamefont {X.-L.}\ \bibnamefont {Qi}}, \bibinfo {author} {\bibfnamefont {H.}~\bibnamefont {Zhang}}, \bibinfo {author} {\bibfnamefont {X.}~\bibnamefont {Dai}}, \bibinfo {author} {\bibfnamefont {Z.}~\bibnamefont {Fang}},\ and\ \bibinfo {author} {\bibfnamefont {S.-C.}\ \bibnamefont {Zhang}},\ }\href {https://doi.org/10.1103/PhysRevB.82.045122} {\bibfield  {journal} {\bibinfo  {journal} {Phys. Rev. B}\ }\textbf {\bibinfo {volume} {82}},\ \bibinfo {pages} {045122} (\bibinfo {year} {2010}{\natexlab{a}})}\BibitemShut {NoStop}%
\bibitem [{\citenamefont {Nechaev}\ and\ \citenamefont {Krasovskii}(2016)}]{nechaev2016}%
  \BibitemOpen
  \bibfield  {author} {\bibinfo {author} {\bibfnamefont {I.~A.}\ \bibnamefont {Nechaev}}\ and\ \bibinfo {author} {\bibfnamefont {E.~E.}\ \bibnamefont {Krasovskii}},\ }\href {https://doi.org/10.1103/PhysRevB.94.201410} {\bibfield  {journal} {\bibinfo  {journal} {Phys. Rev. B}\ }\textbf {\bibinfo {volume} {94}},\ \bibinfo {pages} {201410} (\bibinfo {year} {2016})}\BibitemShut {NoStop}%
\bibitem [{\citenamefont {Bardarson}\ \emph {et~al.}(2010)\citenamefont {Bardarson}, \citenamefont {Brouwer},\ and\ \citenamefont {Moore}}]{Bardarson2010}%
  \BibitemOpen
  \bibfield  {author} {\bibinfo {author} {\bibfnamefont {J.~H.}\ \bibnamefont {Bardarson}}, \bibinfo {author} {\bibfnamefont {P.~W.}\ \bibnamefont {Brouwer}},\ and\ \bibinfo {author} {\bibfnamefont {J.~E.}\ \bibnamefont {Moore}},\ }\href {https://doi.org/10.1103/PhysRevLett.105.156803} {\bibfield  {journal} {\bibinfo  {journal} {Phys. Rev. Lett.}\ }\textbf {\bibinfo {volume} {105}},\ \bibinfo {pages} {156803} (\bibinfo {year} {2010})}\BibitemShut {NoStop}%
\bibitem [{\citenamefont {Brey}\ and\ \citenamefont {Fertig}(2014)}]{brey2014}%
  \BibitemOpen
  \bibfield  {author} {\bibinfo {author} {\bibfnamefont {L.}~\bibnamefont {Brey}}\ and\ \bibinfo {author} {\bibfnamefont {H.~A.}\ \bibnamefont {Fertig}},\ }\href {https://doi.org/10.1103/PhysRevB.89.085305} {\bibfield  {journal} {\bibinfo  {journal} {Phys. Rev. B}\ }\textbf {\bibinfo {volume} {89}},\ \bibinfo {pages} {085305} (\bibinfo {year} {2014})}\BibitemShut {NoStop}%
\bibitem [{\citenamefont {Bardarson}\ and\ \citenamefont {Ilan}(2018)}]{Bardarson_2018}%
  \BibitemOpen
  \bibfield  {author} {\bibinfo {author} {\bibfnamefont {J.~H.}\ \bibnamefont {Bardarson}}\ and\ \bibinfo {author} {\bibfnamefont {R.}~\bibnamefont {Ilan}},\ }\bibinfo {title} {Transport in topological insulator nanowires},\ in\ \href {https://doi.org/10.1007/978-3-319-76388-0_4} {\emph {\bibinfo {booktitle} {Springer Series in Solid-State Sciences}}}\ (\bibinfo  {publisher} {Springer International Publishing},\ \bibinfo {year} {2018})\ p.\ \bibinfo {pages} {93–114}\BibitemShut {NoStop}%
\bibitem [{\citenamefont {Moors}\ \emph {et~al.}(2018)\citenamefont {Moors}, \citenamefont {Sch\"uffelgen}, \citenamefont {Rosenbach}, \citenamefont {Schmitt}, \citenamefont {Sch\"apers},\ and\ \citenamefont {Schmidt}}]{moors2018}%
  \BibitemOpen
  \bibfield  {author} {\bibinfo {author} {\bibfnamefont {K.}~\bibnamefont {Moors}}, \bibinfo {author} {\bibfnamefont {P.}~\bibnamefont {Sch\"uffelgen}}, \bibinfo {author} {\bibfnamefont {D.}~\bibnamefont {Rosenbach}}, \bibinfo {author} {\bibfnamefont {T.}~\bibnamefont {Schmitt}}, \bibinfo {author} {\bibfnamefont {T.}~\bibnamefont {Sch\"apers}},\ and\ \bibinfo {author} {\bibfnamefont {T.~L.}\ \bibnamefont {Schmidt}},\ }\href {https://doi.org/10.1103/PhysRevB.97.245429} {\bibfield  {journal} {\bibinfo  {journal} {Phys. Rev. B}\ }\textbf {\bibinfo {volume} {97}},\ \bibinfo {pages} {245429} (\bibinfo {year} {2018})}\BibitemShut {NoStop}%
\bibitem [{\citenamefont {Ziegler}\ \emph {et~al.}(2018)\citenamefont {Ziegler}, \citenamefont {Kozlovsky}, \citenamefont {Gorini}, \citenamefont {Liu}, \citenamefont {Weish\"aupl}, \citenamefont {Maier}, \citenamefont {Fischer}, \citenamefont {Kozlov}, \citenamefont {Kvon}, \citenamefont {Mikhailov}, \citenamefont {Dvoretsky}, \citenamefont {Richter},\ and\ \citenamefont {Weiss}}]{ziegler_2018}%
  \BibitemOpen
  \bibfield  {author} {\bibinfo {author} {\bibfnamefont {J.}~\bibnamefont {Ziegler}}, \bibinfo {author} {\bibfnamefont {R.}~\bibnamefont {Kozlovsky}}, \bibinfo {author} {\bibfnamefont {C.}~\bibnamefont {Gorini}}, \bibinfo {author} {\bibfnamefont {M.-H.}\ \bibnamefont {Liu}}, \bibinfo {author} {\bibfnamefont {S.}~\bibnamefont {Weish\"aupl}}, \bibinfo {author} {\bibfnamefont {H.}~\bibnamefont {Maier}}, \bibinfo {author} {\bibfnamefont {R.}~\bibnamefont {Fischer}}, \bibinfo {author} {\bibfnamefont {D.~A.}\ \bibnamefont {Kozlov}}, \bibinfo {author} {\bibfnamefont {Z.~D.}\ \bibnamefont {Kvon}}, \bibinfo {author} {\bibfnamefont {N.}~\bibnamefont {Mikhailov}}, \bibinfo {author} {\bibfnamefont {S.~A.}\ \bibnamefont {Dvoretsky}}, \bibinfo {author} {\bibfnamefont {K.}~\bibnamefont {Richter}},\ and\ \bibinfo {author} {\bibfnamefont {D.}~\bibnamefont {Weiss}},\ }\href {https://doi.org/10.1103/PhysRevB.97.035157} {\bibfield  {journal} {\bibinfo  {journal} {Phys. Rev. B}\ }\textbf {\bibinfo {volume} {97}},\ \bibinfo
  {pages} {035157} (\bibinfo {year} {2018})}\BibitemShut {NoStop}%
\bibitem [{\citenamefont {Linder}\ \emph {et~al.}(2009)\citenamefont {Linder}, \citenamefont {Yokoyama},\ and\ \citenamefont {Sudb\o{}}}]{Linder2009}%
  \BibitemOpen
  \bibfield  {author} {\bibinfo {author} {\bibfnamefont {J.}~\bibnamefont {Linder}}, \bibinfo {author} {\bibfnamefont {T.}~\bibnamefont {Yokoyama}},\ and\ \bibinfo {author} {\bibfnamefont {A.}~\bibnamefont {Sudb\o{}}},\ }\href {https://doi.org/10.1103/PhysRevB.80.205401} {\bibfield  {journal} {\bibinfo  {journal} {Phys. Rev. B}\ }\textbf {\bibinfo {volume} {80}},\ \bibinfo {pages} {205401} (\bibinfo {year} {2009})}\BibitemShut {NoStop}%
\bibitem [{\citenamefont {Liu}\ \emph {et~al.}(2010{\natexlab{b}})\citenamefont {Liu}, \citenamefont {Zhang}, \citenamefont {Yan}, \citenamefont {Qi}, \citenamefont {Frauenheim}, \citenamefont {Dai}, \citenamefont {Fang},\ and\ \citenamefont {Zhang}}]{Liu2010A}%
  \BibitemOpen
  \bibfield  {author} {\bibinfo {author} {\bibfnamefont {C.-X.}\ \bibnamefont {Liu}}, \bibinfo {author} {\bibfnamefont {H.}~\bibnamefont {Zhang}}, \bibinfo {author} {\bibfnamefont {B.}~\bibnamefont {Yan}}, \bibinfo {author} {\bibfnamefont {X.-L.}\ \bibnamefont {Qi}}, \bibinfo {author} {\bibfnamefont {T.}~\bibnamefont {Frauenheim}}, \bibinfo {author} {\bibfnamefont {X.}~\bibnamefont {Dai}}, \bibinfo {author} {\bibfnamefont {Z.}~\bibnamefont {Fang}},\ and\ \bibinfo {author} {\bibfnamefont {S.-C.}\ \bibnamefont {Zhang}},\ }\href {https://doi.org/10.1103/PhysRevB.81.041307} {\bibfield  {journal} {\bibinfo  {journal} {Phys. Rev. B}\ }\textbf {\bibinfo {volume} {81}},\ \bibinfo {pages} {041307(R)} (\bibinfo {year} {2010}{\natexlab{b}})}\BibitemShut {NoStop}%
\bibitem [{\citenamefont {Shan}\ \emph {et~al.}(2010)\citenamefont {Shan}, \citenamefont {Lu},\ and\ \citenamefont {Shen}}]{Shan_2010}%
  \BibitemOpen
  \bibfield  {author} {\bibinfo {author} {\bibfnamefont {W.-Y.}\ \bibnamefont {Shan}}, \bibinfo {author} {\bibfnamefont {H.-Z.}\ \bibnamefont {Lu}},\ and\ \bibinfo {author} {\bibfnamefont {S.-Q.}\ \bibnamefont {Shen}},\ }\href {https://doi.org/10.1088/1367-2630/12/4/043048} {\bibfield  {journal} {\bibinfo  {journal} {New J. Phys.}\ }\textbf {\bibinfo {volume} {12}},\ \bibinfo {pages} {043048} (\bibinfo {year} {2010})}\BibitemShut {NoStop}%
\bibitem [{\citenamefont {Lu}\ \emph {et~al.}(2010)\citenamefont {Lu}, \citenamefont {Shan}, \citenamefont {Yao}, \citenamefont {Niu},\ and\ \citenamefont {Shen}}]{Lu2010}%
  \BibitemOpen
  \bibfield  {author} {\bibinfo {author} {\bibfnamefont {H.-Z.}\ \bibnamefont {Lu}}, \bibinfo {author} {\bibfnamefont {W.-Y.}\ \bibnamefont {Shan}}, \bibinfo {author} {\bibfnamefont {W.}~\bibnamefont {Yao}}, \bibinfo {author} {\bibfnamefont {Q.}~\bibnamefont {Niu}},\ and\ \bibinfo {author} {\bibfnamefont {S.-Q.}\ \bibnamefont {Shen}},\ }\href {https://doi.org/10.1103/PhysRevB.81.115407} {\bibfield  {journal} {\bibinfo  {journal} {Phys. Rev. B}\ }\textbf {\bibinfo {volume} {81}},\ \bibinfo {pages} {115407} (\bibinfo {year} {2010})}\BibitemShut {NoStop}%
\bibitem [{\citenamefont {Zhang}\ \emph {et~al.}(2015)\citenamefont {Zhang}, \citenamefont {Lu},\ and\ \citenamefont {Shen}}]{Zhang_2015}%
  \BibitemOpen
  \bibfield  {author} {\bibinfo {author} {\bibfnamefont {S.-B.}\ \bibnamefont {Zhang}}, \bibinfo {author} {\bibfnamefont {H.-Z.}\ \bibnamefont {Lu}},\ and\ \bibinfo {author} {\bibfnamefont {S.-Q.}\ \bibnamefont {Shen}},\ }\href {https://doi.org/10.1038/srep13277} {\bibfield  {journal} {\bibinfo  {journal} {Sci. Rep.}\ }\textbf {\bibinfo {volume} {5}},\ \bibinfo {pages} {13277} (\bibinfo {year} {2015})}\BibitemShut {NoStop}%
\bibitem [{\citenamefont {Leis}\ \emph {et~al.}(2022)\citenamefont {Leis}, \citenamefont {Schleenvoigt}, \citenamefont {Moors}, \citenamefont {Soltner}, \citenamefont {Cherepanov}, \citenamefont {Schüffelgen}, \citenamefont {Mussler}, \citenamefont {Grützmacher}, \citenamefont {Voigtländer}, \citenamefont {Lüpke},\ and\ \citenamefont {Tautz}}]{Leis_2022}%
  \BibitemOpen
  \bibfield  {author} {\bibinfo {author} {\bibfnamefont {A.}~\bibnamefont {Leis}}, \bibinfo {author} {\bibfnamefont {M.}~\bibnamefont {Schleenvoigt}}, \bibinfo {author} {\bibfnamefont {K.}~\bibnamefont {Moors}}, \bibinfo {author} {\bibfnamefont {H.}~\bibnamefont {Soltner}}, \bibinfo {author} {\bibfnamefont {V.}~\bibnamefont {Cherepanov}}, \bibinfo {author} {\bibfnamefont {P.}~\bibnamefont {Schüffelgen}}, \bibinfo {author} {\bibfnamefont {G.}~\bibnamefont {Mussler}}, \bibinfo {author} {\bibfnamefont {D.}~\bibnamefont {Grützmacher}}, \bibinfo {author} {\bibfnamefont {B.}~\bibnamefont {Voigtländer}}, \bibinfo {author} {\bibfnamefont {F.}~\bibnamefont {Lüpke}},\ and\ \bibinfo {author} {\bibfnamefont {F.~S.}\ \bibnamefont {Tautz}},\ }\href {https://doi.org/10.1002/qute.202200043} {\bibfield  {journal} {\bibinfo  {journal} {Adv. Quantum Technol.}\ }\textbf {\bibinfo {volume} {5}},\ \bibinfo {pages} {2200043} (\bibinfo {year} {2022})}\BibitemShut {NoStop}%
\bibitem [{\citenamefont {Yu}\ \emph {et~al.}(2010)\citenamefont {Yu}, \citenamefont {Zhang}, \citenamefont {Zhang}, \citenamefont {Zhang}, \citenamefont {Dai},\ and\ \citenamefont {Fang}}]{Yu2010}%
  \BibitemOpen
  \bibfield  {author} {\bibinfo {author} {\bibfnamefont {R.}~\bibnamefont {Yu}}, \bibinfo {author} {\bibfnamefont {W.}~\bibnamefont {Zhang}}, \bibinfo {author} {\bibfnamefont {H.-J.}\ \bibnamefont {Zhang}}, \bibinfo {author} {\bibfnamefont {S.-C.}\ \bibnamefont {Zhang}}, \bibinfo {author} {\bibfnamefont {X.}~\bibnamefont {Dai}},\ and\ \bibinfo {author} {\bibfnamefont {Z.}~\bibnamefont {Fang}},\ }\href {https://doi.org/10.1126/science.1187485} {\bibfield  {journal} {\bibinfo  {journal} {Science}\ }\textbf {\bibinfo {volume} {329}},\ \bibinfo {pages} {61} (\bibinfo {year} {2010})}\BibitemShut {NoStop}%
\bibitem [{\citenamefont {Chang}\ \emph {et~al.}(2013)\citenamefont {Chang}, \citenamefont {Zhang}, \citenamefont {Feng}, \citenamefont {Shen}, \citenamefont {Zhang}, \citenamefont {Guo}, \citenamefont {Li}, \citenamefont {Ou}, \citenamefont {Wei}, \citenamefont {Wang}, \citenamefont {Ji}, \citenamefont {Feng}, \citenamefont {Ji}, \citenamefont {Chen}, \citenamefont {Jia}, \citenamefont {Dai}, \citenamefont {Fang}, \citenamefont {Zhang}, \citenamefont {He}, \citenamefont {Wang}, \citenamefont {Lu}, \citenamefont {Ma},\ and\ \citenamefont {Xue}}]{Chang2013}%
  \BibitemOpen
  \bibfield  {author} {\bibinfo {author} {\bibfnamefont {C.-Z.}\ \bibnamefont {Chang}}, \bibinfo {author} {\bibfnamefont {J.}~\bibnamefont {Zhang}}, \bibinfo {author} {\bibfnamefont {X.}~\bibnamefont {Feng}}, \bibinfo {author} {\bibfnamefont {J.}~\bibnamefont {Shen}}, \bibinfo {author} {\bibfnamefont {Z.}~\bibnamefont {Zhang}}, \bibinfo {author} {\bibfnamefont {M.}~\bibnamefont {Guo}}, \bibinfo {author} {\bibfnamefont {K.}~\bibnamefont {Li}}, \bibinfo {author} {\bibfnamefont {Y.}~\bibnamefont {Ou}}, \bibinfo {author} {\bibfnamefont {P.}~\bibnamefont {Wei}}, \bibinfo {author} {\bibfnamefont {L.-L.}\ \bibnamefont {Wang}}, \bibinfo {author} {\bibfnamefont {Z.-Q.}\ \bibnamefont {Ji}}, \bibinfo {author} {\bibfnamefont {Y.}~\bibnamefont {Feng}}, \bibinfo {author} {\bibfnamefont {S.}~\bibnamefont {Ji}}, \bibinfo {author} {\bibfnamefont {X.}~\bibnamefont {Chen}}, \bibinfo {author} {\bibfnamefont {J.}~\bibnamefont {Jia}}, \bibinfo {author} {\bibfnamefont {X.}~\bibnamefont {Dai}}, \bibinfo {author} {\bibfnamefont
  {Z.}~\bibnamefont {Fang}}, \bibinfo {author} {\bibfnamefont {S.-C.}\ \bibnamefont {Zhang}}, \bibinfo {author} {\bibfnamefont {K.}~\bibnamefont {He}}, \bibinfo {author} {\bibfnamefont {Y.}~\bibnamefont {Wang}}, \bibinfo {author} {\bibfnamefont {L.}~\bibnamefont {Lu}}, \bibinfo {author} {\bibfnamefont {X.-C.}\ \bibnamefont {Ma}},\ and\ \bibinfo {author} {\bibfnamefont {Q.-K.}\ \bibnamefont {Xue}},\ }\href {https://doi.org/10.1126/science.1234414} {\bibfield  {journal} {\bibinfo  {journal} {Science}\ }\textbf {\bibinfo {volume} {340}},\ \bibinfo {pages} {167} (\bibinfo {year} {2013})}\BibitemShut {NoStop}%
\bibitem [{\citenamefont {Kou}\ \emph {et~al.}(2013)\citenamefont {Kou}, \citenamefont {Lang}, \citenamefont {Fan}, \citenamefont {Jiang}, \citenamefont {Nie}, \citenamefont {Zhang}, \citenamefont {Jiang}, \citenamefont {Wang}, \citenamefont {Yao}, \citenamefont {He},\ and\ \citenamefont {Wang}}]{Kou2013}%
  \BibitemOpen
  \bibfield  {author} {\bibinfo {author} {\bibfnamefont {X.}~\bibnamefont {Kou}}, \bibinfo {author} {\bibfnamefont {M.}~\bibnamefont {Lang}}, \bibinfo {author} {\bibfnamefont {Y.}~\bibnamefont {Fan}}, \bibinfo {author} {\bibfnamefont {Y.}~\bibnamefont {Jiang}}, \bibinfo {author} {\bibfnamefont {T.}~\bibnamefont {Nie}}, \bibinfo {author} {\bibfnamefont {J.}~\bibnamefont {Zhang}}, \bibinfo {author} {\bibfnamefont {W.}~\bibnamefont {Jiang}}, \bibinfo {author} {\bibfnamefont {Y.}~\bibnamefont {Wang}}, \bibinfo {author} {\bibfnamefont {Y.}~\bibnamefont {Yao}}, \bibinfo {author} {\bibfnamefont {L.}~\bibnamefont {He}},\ and\ \bibinfo {author} {\bibfnamefont {K.~L.}\ \bibnamefont {Wang}},\ }\href {https://doi.org/10.1021/nn4038145} {\bibfield  {journal} {\bibinfo  {journal} {ACS Nano}\ }\textbf {\bibinfo {volume} {7}},\ \bibinfo {pages} {9205} (\bibinfo {year} {2013})}\BibitemShut {NoStop}%
\bibitem [{\citenamefont {Wang}\ \emph {et~al.}(2015{\natexlab{a}})\citenamefont {Wang}, \citenamefont {Lian},\ and\ \citenamefont {Zhang}}]{Wang_2015}%
  \BibitemOpen
  \bibfield  {author} {\bibinfo {author} {\bibfnamefont {J.}~\bibnamefont {Wang}}, \bibinfo {author} {\bibfnamefont {B.}~\bibnamefont {Lian}},\ and\ \bibinfo {author} {\bibfnamefont {S.-C.}\ \bibnamefont {Zhang}},\ }\href {https://doi.org/10.1088/0031-8949/2015/T164/014003} {\bibfield  {journal} {\bibinfo  {journal} {Phys. Scripta}\ }\textbf {\bibinfo {volume} {2015}},\ \bibinfo {pages} {014003} (\bibinfo {year} {2015}{\natexlab{a}})}\BibitemShut {NoStop}%
\bibitem [{\citenamefont {Yasuda}\ \emph {et~al.}(2017)\citenamefont {Yasuda}, \citenamefont {Mogi}, \citenamefont {Yoshimi}, \citenamefont {Tsukazaki}, \citenamefont {Takahashi}, \citenamefont {Kawasaki}, \citenamefont {Kagawa},\ and\ \citenamefont {Tokura}}]{Yasuda2017}%
  \BibitemOpen
  \bibfield  {author} {\bibinfo {author} {\bibfnamefont {K.}~\bibnamefont {Yasuda}}, \bibinfo {author} {\bibfnamefont {M.}~\bibnamefont {Mogi}}, \bibinfo {author} {\bibfnamefont {R.}~\bibnamefont {Yoshimi}}, \bibinfo {author} {\bibfnamefont {A.}~\bibnamefont {Tsukazaki}}, \bibinfo {author} {\bibfnamefont {K.~S.}\ \bibnamefont {Takahashi}}, \bibinfo {author} {\bibfnamefont {M.}~\bibnamefont {Kawasaki}}, \bibinfo {author} {\bibfnamefont {F.}~\bibnamefont {Kagawa}},\ and\ \bibinfo {author} {\bibfnamefont {Y.}~\bibnamefont {Tokura}},\ }\href {https://doi.org/10.1126/science.aan5991} {\bibfield  {journal} {\bibinfo  {journal} {Science}\ }\textbf {\bibinfo {volume} {358}},\ \bibinfo {pages} {1311} (\bibinfo {year} {2017})}\BibitemShut {NoStop}%
\bibitem [{\citenamefont {Sun}\ \emph {et~al.}(2019)\citenamefont {Sun}, \citenamefont {Xia}, \citenamefont {Chen}, \citenamefont {Zhang}, \citenamefont {Liu}, \citenamefont {Yao}, \citenamefont {Tang}, \citenamefont {Zhao}, \citenamefont {Xu},\ and\ \citenamefont {Liu}}]{Sun_2019}%
  \BibitemOpen
  \bibfield  {author} {\bibinfo {author} {\bibfnamefont {H.}~\bibnamefont {Sun}}, \bibinfo {author} {\bibfnamefont {B.}~\bibnamefont {Xia}}, \bibinfo {author} {\bibfnamefont {Z.}~\bibnamefont {Chen}}, \bibinfo {author} {\bibfnamefont {Y.}~\bibnamefont {Zhang}}, \bibinfo {author} {\bibfnamefont {P.}~\bibnamefont {Liu}}, \bibinfo {author} {\bibfnamefont {Q.}~\bibnamefont {Yao}}, \bibinfo {author} {\bibfnamefont {H.}~\bibnamefont {Tang}}, \bibinfo {author} {\bibfnamefont {Y.}~\bibnamefont {Zhao}}, \bibinfo {author} {\bibfnamefont {H.}~\bibnamefont {Xu}},\ and\ \bibinfo {author} {\bibfnamefont {Q.}~\bibnamefont {Liu}},\ }\href {https://doi.org/10.1103/PhysRevLett.123.096401} {\bibfield  {journal} {\bibinfo  {journal} {Phys. Rev. Lett.}\ }\textbf {\bibinfo {volume} {123}},\ \bibinfo {pages} {096401} (\bibinfo {year} {2019})}\BibitemShut {NoStop}%
\bibitem [{\citenamefont {Tokura}\ \emph {et~al.}(2019)\citenamefont {Tokura}, \citenamefont {Yasuda},\ and\ \citenamefont {Tsukazaki}}]{Tokura2019}%
  \BibitemOpen
  \bibfield  {author} {\bibinfo {author} {\bibfnamefont {Y.}~\bibnamefont {Tokura}}, \bibinfo {author} {\bibfnamefont {K.}~\bibnamefont {Yasuda}},\ and\ \bibinfo {author} {\bibfnamefont {A.}~\bibnamefont {Tsukazaki}},\ }\href {https://doi.org/10.1038/s42254-018-0011-5} {\bibfield  {journal} {\bibinfo  {journal} {Nat. Rev. Phys.}\ }\textbf {\bibinfo {volume} {1}},\ \bibinfo {pages} {126} (\bibinfo {year} {2019})}\BibitemShut {NoStop}%
\bibitem [{\citenamefont {Yasuda}\ \emph {et~al.}(2020)\citenamefont {Yasuda}, \citenamefont {Morimoto}, \citenamefont {Yoshimi}, \citenamefont {Mogi}, \citenamefont {Tsukazaki}, \citenamefont {Kawamura}, \citenamefont {Takahashi}, \citenamefont {Kawasaki}, \citenamefont {Nagaosa},\ and\ \citenamefont {Tokura}}]{Yasuda2020}%
  \BibitemOpen
  \bibfield  {author} {\bibinfo {author} {\bibfnamefont {K.}~\bibnamefont {Yasuda}}, \bibinfo {author} {\bibfnamefont {T.}~\bibnamefont {Morimoto}}, \bibinfo {author} {\bibfnamefont {R.}~\bibnamefont {Yoshimi}}, \bibinfo {author} {\bibfnamefont {M.}~\bibnamefont {Mogi}}, \bibinfo {author} {\bibfnamefont {A.}~\bibnamefont {Tsukazaki}}, \bibinfo {author} {\bibfnamefont {M.}~\bibnamefont {Kawamura}}, \bibinfo {author} {\bibfnamefont {K.}~\bibnamefont {Takahashi}}, \bibinfo {author} {\bibfnamefont {M.}~\bibnamefont {Kawasaki}}, \bibinfo {author} {\bibfnamefont {N.}~\bibnamefont {Nagaosa}},\ and\ \bibinfo {author} {\bibfnamefont {Y.}~\bibnamefont {Tokura}},\ }\href {https://doi.org/10.1038/s41565-020-0733-2} {\bibfield  {journal} {\bibinfo  {journal} {Nat. Nanotechnol.}\ }\textbf {\bibinfo {volume} {15}},\ \bibinfo {pages} {831–835} (\bibinfo {year} {2020})}\BibitemShut {NoStop}%
\bibitem [{\citenamefont {Wang}\ \emph {et~al.}(2021)\citenamefont {Wang}, \citenamefont {Ge}, \citenamefont {Li}, \citenamefont {Liu}, \citenamefont {Xu},\ and\ \citenamefont {Wang}}]{Wang2021}%
  \BibitemOpen
  \bibfield  {author} {\bibinfo {author} {\bibfnamefont {P.}~\bibnamefont {Wang}}, \bibinfo {author} {\bibfnamefont {J.}~\bibnamefont {Ge}}, \bibinfo {author} {\bibfnamefont {J.}~\bibnamefont {Li}}, \bibinfo {author} {\bibfnamefont {Y.}~\bibnamefont {Liu}}, \bibinfo {author} {\bibfnamefont {Y.}~\bibnamefont {Xu}},\ and\ \bibinfo {author} {\bibfnamefont {J.}~\bibnamefont {Wang}},\ }\href {https://doi.org/https://doi.org/10.1016/j.xinn.2021.100098} {\bibfield  {journal} {\bibinfo  {journal} {Innovation}\ }\textbf {\bibinfo {volume} {2}},\ \bibinfo {pages} {100098} (\bibinfo {year} {2021})}\BibitemShut {NoStop}%
\bibitem [{\citenamefont {Qiu}\ \emph {et~al.}(2022)\citenamefont {Qiu}, \citenamefont {Zhang}, \citenamefont {Deng}, \citenamefont {Chong}, \citenamefont {Tai}, \citenamefont {Eckberg},\ and\ \citenamefont {Wang}}]{Qiu_2022}%
  \BibitemOpen
  \bibfield  {author} {\bibinfo {author} {\bibfnamefont {G.}~\bibnamefont {Qiu}}, \bibinfo {author} {\bibfnamefont {P.}~\bibnamefont {Zhang}}, \bibinfo {author} {\bibfnamefont {P.}~\bibnamefont {Deng}}, \bibinfo {author} {\bibfnamefont {S.~K.}\ \bibnamefont {Chong}}, \bibinfo {author} {\bibfnamefont {L.}~\bibnamefont {Tai}}, \bibinfo {author} {\bibfnamefont {C.}~\bibnamefont {Eckberg}},\ and\ \bibinfo {author} {\bibfnamefont {K.~L.}\ \bibnamefont {Wang}},\ }\href {https://doi.org/10.1103/PhysRevLett.128.217704} {\bibfield  {journal} {\bibinfo  {journal} {Phys. Rev. Lett.}\ }\textbf {\bibinfo {volume} {128}},\ \bibinfo {pages} {217704} (\bibinfo {year} {2022})}\BibitemShut {NoStop}%
\bibitem [{\citenamefont {Atanov}\ \emph {et~al.}(2024)\citenamefont {Atanov}, \citenamefont {Tai}, \citenamefont {Xie}, \citenamefont {Ng}, \citenamefont {Hammond}, \citenamefont {{Manfred Ho}}, \citenamefont {Koo}, \citenamefont {Li}, \citenamefont {Ho}, \citenamefont {Lyu}, \citenamefont {Chong}, \citenamefont {Zhang}, \citenamefont {Tai}, \citenamefont {Wang}, \citenamefont {Law}, \citenamefont {Wang},\ and\ \citenamefont {Lortz}}]{Atanov2024}%
  \BibitemOpen
  \bibfield  {author} {\bibinfo {author} {\bibfnamefont {O.}~\bibnamefont {Atanov}}, \bibinfo {author} {\bibfnamefont {W.~T.}\ \bibnamefont {Tai}}, \bibinfo {author} {\bibfnamefont {Y.-M.}\ \bibnamefont {Xie}}, \bibinfo {author} {\bibfnamefont {Y.~H.}\ \bibnamefont {Ng}}, \bibinfo {author} {\bibfnamefont {M.~A.}\ \bibnamefont {Hammond}}, \bibinfo {author} {\bibfnamefont {T.~S.}\ \bibnamefont {{Manfred Ho}}}, \bibinfo {author} {\bibfnamefont {T.~H.}\ \bibnamefont {Koo}}, \bibinfo {author} {\bibfnamefont {H.}~\bibnamefont {Li}}, \bibinfo {author} {\bibfnamefont {S.~L.}\ \bibnamefont {Ho}}, \bibinfo {author} {\bibfnamefont {J.}~\bibnamefont {Lyu}}, \bibinfo {author} {\bibfnamefont {S.}~\bibnamefont {Chong}}, \bibinfo {author} {\bibfnamefont {P.}~\bibnamefont {Zhang}}, \bibinfo {author} {\bibfnamefont {L.}~\bibnamefont {Tai}}, \bibinfo {author} {\bibfnamefont {J.}~\bibnamefont {Wang}}, \bibinfo {author} {\bibfnamefont {K.~T.}\ \bibnamefont {Law}}, \bibinfo {author} {\bibfnamefont {K.~L.}\ \bibnamefont {Wang}},\
  and\ \bibinfo {author} {\bibfnamefont {R.}~\bibnamefont {Lortz}},\ }\href {https://doi.org/https://doi.org/10.1016/j.xcrp.2023.101762} {\bibfield  {journal} {\bibinfo  {journal} {Cell Rep. Phys. Sci.}\ }\textbf {\bibinfo {volume} {5}},\ \bibinfo {pages} {101762} (\bibinfo {year} {2024})}\BibitemShut {NoStop}%
\bibitem [{\citenamefont {Wang}\ \emph {et~al.}(2015{\natexlab{b}})\citenamefont {Wang}, \citenamefont {Zhou}, \citenamefont {Lian},\ and\ \citenamefont {Zhang}}]{Wang2015_chiral}%
  \BibitemOpen
  \bibfield  {author} {\bibinfo {author} {\bibfnamefont {J.}~\bibnamefont {Wang}}, \bibinfo {author} {\bibfnamefont {Q.}~\bibnamefont {Zhou}}, \bibinfo {author} {\bibfnamefont {B.}~\bibnamefont {Lian}},\ and\ \bibinfo {author} {\bibfnamefont {S.-C.}\ \bibnamefont {Zhang}},\ }\href {https://doi.org/10.1103/PhysRevB.92.064520} {\bibfield  {journal} {\bibinfo  {journal} {Phys. Rev. B}\ }\textbf {\bibinfo {volume} {92}},\ \bibinfo {pages} {064520} (\bibinfo {year} {2015}{\natexlab{b}})}\BibitemShut {NoStop}%
\bibitem [{\citenamefont {Zeng}\ \emph {et~al.}(2018)\citenamefont {Zeng}, \citenamefont {Lei}, \citenamefont {Chaudhary},\ and\ \citenamefont {MacDonald}}]{Zeng2018}%
  \BibitemOpen
  \bibfield  {author} {\bibinfo {author} {\bibfnamefont {Y.}~\bibnamefont {Zeng}}, \bibinfo {author} {\bibfnamefont {C.}~\bibnamefont {Lei}}, \bibinfo {author} {\bibfnamefont {G.}~\bibnamefont {Chaudhary}},\ and\ \bibinfo {author} {\bibfnamefont {A.~H.}\ \bibnamefont {MacDonald}},\ }\href {https://doi.org/10.1103/PhysRevB.97.081102} {\bibfield  {journal} {\bibinfo  {journal} {Phys. Rev. B}\ }\textbf {\bibinfo {volume} {97}},\ \bibinfo {pages} {081102(R)} (\bibinfo {year} {2018})}\BibitemShut {NoStop}%
\bibitem [{\citenamefont {Chen}\ \emph {et~al.}(2018)\citenamefont {Chen}, \citenamefont {Xie}, \citenamefont {Liu}, \citenamefont {Lee},\ and\ \citenamefont {Law}}]{Chen2018}%
  \BibitemOpen
  \bibfield  {author} {\bibinfo {author} {\bibfnamefont {C.-Z.}\ \bibnamefont {Chen}}, \bibinfo {author} {\bibfnamefont {Y.-M.}\ \bibnamefont {Xie}}, \bibinfo {author} {\bibfnamefont {J.}~\bibnamefont {Liu}}, \bibinfo {author} {\bibfnamefont {P.~A.}\ \bibnamefont {Lee}},\ and\ \bibinfo {author} {\bibfnamefont {K.~T.}\ \bibnamefont {Law}},\ }\href {https://doi.org/10.1103/PhysRevB.97.104504} {\bibfield  {journal} {\bibinfo  {journal} {Phys. Rev. B}\ }\textbf {\bibinfo {volume} {97}},\ \bibinfo {pages} {104504} (\bibinfo {year} {2018})}\BibitemShut {NoStop}%
\bibitem [{\citenamefont {Mandal}\ \emph {et~al.}(2022)\citenamefont {Mandal}, \citenamefont {Taufertshöfer}, \citenamefont {Lunczer}, \citenamefont {Stehno}, \citenamefont {Gould},\ and\ \citenamefont {Molenkamp}}]{Mandal2022}%
  \BibitemOpen
  \bibfield  {author} {\bibinfo {author} {\bibfnamefont {P.}~\bibnamefont {Mandal}}, \bibinfo {author} {\bibfnamefont {N.}~\bibnamefont {Taufertshöfer}}, \bibinfo {author} {\bibfnamefont {L.}~\bibnamefont {Lunczer}}, \bibinfo {author} {\bibfnamefont {M.~P.}\ \bibnamefont {Stehno}}, \bibinfo {author} {\bibfnamefont {C.}~\bibnamefont {Gould}},\ and\ \bibinfo {author} {\bibfnamefont {L.~W.}\ \bibnamefont {Molenkamp}},\ }\href {https://doi.org/10.1021/acs.nanolett.1c04903} {\bibfield  {journal} {\bibinfo  {journal} {Nano Lett.}\ }\textbf {\bibinfo {volume} {22}},\ \bibinfo {pages} {3557} (\bibinfo {year} {2022})}\BibitemShut {NoStop}%
\bibitem [{\citenamefont {Uday}\ \emph {et~al.}(2023)\citenamefont {Uday}, \citenamefont {Lippertz}, \citenamefont {Moors}, \citenamefont {Legg}, \citenamefont {Bliesener}, \citenamefont {Pereira}, \citenamefont {Taskin},\ and\ \citenamefont {Ando}}]{uday2023induced}%
  \BibitemOpen
  \bibfield  {author} {\bibinfo {author} {\bibfnamefont {A.}~\bibnamefont {Uday}}, \bibinfo {author} {\bibfnamefont {G.}~\bibnamefont {Lippertz}}, \bibinfo {author} {\bibfnamefont {K.}~\bibnamefont {Moors}}, \bibinfo {author} {\bibfnamefont {H.~F.}\ \bibnamefont {Legg}}, \bibinfo {author} {\bibfnamefont {A.}~\bibnamefont {Bliesener}}, \bibinfo {author} {\bibfnamefont {L.~M.~C.}\ \bibnamefont {Pereira}}, \bibinfo {author} {\bibfnamefont {A.~A.}\ \bibnamefont {Taskin}},\ and\ \bibinfo {author} {\bibfnamefont {Y.}~\bibnamefont {Ando}}\ }\href {https://doi.org/10.48550/arXiv.2307.08578} {10.48550/arXiv.2307.08578} (\bibinfo {year} {2023}),\ \Eprint {https://arxiv.org/abs/2307.08578} {arXiv:2307.08578 [cond-mat.mes-hall]} \BibitemShut {NoStop}%
\bibitem [{\citenamefont {F\"orster}\ \emph {et~al.}(2015)\citenamefont {F\"orster}, \citenamefont {Kr\"uger},\ and\ \citenamefont {Rohlfing}}]{Forster2015}%
  \BibitemOpen
  \bibfield  {author} {\bibinfo {author} {\bibfnamefont {T.}~\bibnamefont {F\"orster}}, \bibinfo {author} {\bibfnamefont {P.}~\bibnamefont {Kr\"uger}},\ and\ \bibinfo {author} {\bibfnamefont {M.}~\bibnamefont {Rohlfing}},\ }\href {https://doi.org/10.1103/PhysRevB.92.201404} {\bibfield  {journal} {\bibinfo  {journal} {Phys. Rev. B}\ }\textbf {\bibinfo {volume} {92}},\ \bibinfo {pages} {201404} (\bibinfo {year} {2015})}\BibitemShut {NoStop}%
\bibitem [{\citenamefont {F\"orster}\ \emph {et~al.}(2016)\citenamefont {F\"orster}, \citenamefont {Kr\"uger},\ and\ \citenamefont {Rohlfing}}]{Forster2016}%
  \BibitemOpen
  \bibfield  {author} {\bibinfo {author} {\bibfnamefont {T.}~\bibnamefont {F\"orster}}, \bibinfo {author} {\bibfnamefont {P.}~\bibnamefont {Kr\"uger}},\ and\ \bibinfo {author} {\bibfnamefont {M.}~\bibnamefont {Rohlfing}},\ }\href {https://doi.org/10.1103/PhysRevB.93.205442} {\bibfield  {journal} {\bibinfo  {journal} {Phys. Rev. B}\ }\textbf {\bibinfo {volume} {93}},\ \bibinfo {pages} {205442} (\bibinfo {year} {2016})}\BibitemShut {NoStop}%
\bibitem [{\citenamefont {Wang}\ \emph {et~al.}(2019)\citenamefont {Wang}, \citenamefont {Zhou}, \citenamefont {Jiang}, \citenamefont {Sun}, \citenamefont {Zang}, \citenamefont {Gong}, \citenamefont {Zhang}, \citenamefont {Tong}, \citenamefont {Xie}, \citenamefont {Liu}, \citenamefont {Chen}, \citenamefont {He},\ and\ \citenamefont {Xue}}]{Wang2019}%
  \BibitemOpen
  \bibfield  {author} {\bibinfo {author} {\bibfnamefont {Z.}~\bibnamefont {Wang}}, \bibinfo {author} {\bibfnamefont {T.}~\bibnamefont {Zhou}}, \bibinfo {author} {\bibfnamefont {T.}~\bibnamefont {Jiang}}, \bibinfo {author} {\bibfnamefont {H.}~\bibnamefont {Sun}}, \bibinfo {author} {\bibfnamefont {Y.}~\bibnamefont {Zang}}, \bibinfo {author} {\bibfnamefont {Y.}~\bibnamefont {Gong}}, \bibinfo {author} {\bibfnamefont {J.}~\bibnamefont {Zhang}}, \bibinfo {author} {\bibfnamefont {M.}~\bibnamefont {Tong}}, \bibinfo {author} {\bibfnamefont {X.}~\bibnamefont {Xie}}, \bibinfo {author} {\bibfnamefont {Q.}~\bibnamefont {Liu}}, \bibinfo {author} {\bibfnamefont {C.}~\bibnamefont {Chen}}, \bibinfo {author} {\bibfnamefont {K.}~\bibnamefont {He}},\ and\ \bibinfo {author} {\bibfnamefont {Q.-K.}\ \bibnamefont {Xue}},\ }\href {https://doi.org/10.1021/acs.nanolett.9b01641} {\bibfield  {journal} {\bibinfo  {journal} {Nano Lett.}\ }\textbf {\bibinfo {volume} {19}},\ \bibinfo {pages} {4627} (\bibinfo {year} {2019})}\BibitemShut
  {NoStop}%
\bibitem [{\citenamefont {Liu}\ \emph {et~al.}(2023)\citenamefont {Liu}, \citenamefont {Yang}, \citenamefont {Xue}, \citenamefont {Gai}, \citenamefont {Sun}, \citenamefont {Li}, \citenamefont {Gong}, \citenamefont {Li}, \citenamefont {Xie}, \citenamefont {He}, \citenamefont {Zhang}, \citenamefont {Xue},\ and\ \citenamefont {Cheng}}]{Liu2023}%
  \BibitemOpen
  \bibfield  {author} {\bibinfo {author} {\bibfnamefont {J.-n.}\ \bibnamefont {Liu}}, \bibinfo {author} {\bibfnamefont {X.}~\bibnamefont {Yang}}, \bibinfo {author} {\bibfnamefont {H.}~\bibnamefont {Xue}}, \bibinfo {author} {\bibfnamefont {X.-s.}\ \bibnamefont {Gai}}, \bibinfo {author} {\bibfnamefont {R.}~\bibnamefont {Sun}}, \bibinfo {author} {\bibfnamefont {Y.}~\bibnamefont {Li}}, \bibinfo {author} {\bibfnamefont {Z.-z.}\ \bibnamefont {Gong}}, \bibinfo {author} {\bibfnamefont {N.}~\bibnamefont {Li}}, \bibinfo {author} {\bibfnamefont {Z.-K.}\ \bibnamefont {Xie}}, \bibinfo {author} {\bibfnamefont {W.}~\bibnamefont {He}}, \bibinfo {author} {\bibfnamefont {X.-Q.}\ \bibnamefont {Zhang}}, \bibinfo {author} {\bibfnamefont {D.}~\bibnamefont {Xue}},\ and\ \bibinfo {author} {\bibfnamefont {Z.-H.}\ \bibnamefont {Cheng}},\ }\href {https://doi.org/10.1038/s41467-023-40035-0} {\bibfield  {journal} {\bibinfo  {journal} {Nat. Commun.}\ }\textbf {\bibinfo {volume} {14}},\ \bibinfo {pages} {4424} (\bibinfo {year}
  {2023})}\BibitemShut {NoStop}%
\bibitem [{\citenamefont {Shen}(2012)}]{shen_book}%
  \BibitemOpen
  \bibfield  {author} {\bibinfo {author} {\bibfnamefont {S.-Q.}\ \bibnamefont {Shen}},\ }\href {https://doi.org/10.1007/978-3-642-32858-9} {\emph {\bibinfo {title} {Topological Insulators: Dirac Equation in Condensed Matters}}}\ (\bibinfo  {publisher} {Springer Berlin Heidelberg},\ \bibinfo {year} {2012})\BibitemShut {NoStop}%
\bibitem [{\citenamefont {Nielsen}\ and\ \citenamefont {Ninomiya}(1981)}]{NIELSEN198120}%
  \BibitemOpen
  \bibfield  {author} {\bibinfo {author} {\bibfnamefont {H.}~\bibnamefont {Nielsen}}\ and\ \bibinfo {author} {\bibfnamefont {M.}~\bibnamefont {Ninomiya}},\ }\href {https://doi.org/https://doi.org/10.1016/0550-3213(81)90361-8} {\bibfield  {journal} {\bibinfo  {journal} {Nucl. Phys. B}\ }\textbf {\bibinfo {volume} {185}},\ \bibinfo {pages} {20} (\bibinfo {year} {1981})}\BibitemShut {NoStop}%
\bibitem [{\citenamefont {Datta}(2005)}]{datta_2005}%
  \BibitemOpen
  \bibfield  {author} {\bibinfo {author} {\bibfnamefont {S.}~\bibnamefont {Datta}},\ }\href {https://doi.org/10.1017/CBO9781139164313} {\emph {\bibinfo {title} {Quantum Transport: Atom to Transistor}}}\ (\bibinfo  {publisher} {Cambridge University Press},\ \bibinfo {year} {2005})\BibitemShut {NoStop}%
\bibitem [{\citenamefont {Zhang}\ \emph {et~al.}(2010)\citenamefont {Zhang}, \citenamefont {He}, \citenamefont {Chang}, \citenamefont {Song}, \citenamefont {Wang}, \citenamefont {Chen}, \citenamefont {Jia}, \citenamefont {Fang}, \citenamefont {Dai}, \citenamefont {Shan}, \citenamefont {Shen}, \citenamefont {Niu}, \citenamefont {Qi}, \citenamefont {Zhang}, \citenamefont {Ma},\ and\ \citenamefont {Xue}}]{Zhang2010A}%
  \BibitemOpen
  \bibfield  {author} {\bibinfo {author} {\bibfnamefont {Y.}~\bibnamefont {Zhang}}, \bibinfo {author} {\bibfnamefont {K.}~\bibnamefont {He}}, \bibinfo {author} {\bibfnamefont {C.-Z.}\ \bibnamefont {Chang}}, \bibinfo {author} {\bibfnamefont {C.-L.}\ \bibnamefont {Song}}, \bibinfo {author} {\bibfnamefont {L.-L.}\ \bibnamefont {Wang}}, \bibinfo {author} {\bibfnamefont {X.}~\bibnamefont {Chen}}, \bibinfo {author} {\bibfnamefont {J.-F.}\ \bibnamefont {Jia}}, \bibinfo {author} {\bibfnamefont {Z.}~\bibnamefont {Fang}}, \bibinfo {author} {\bibfnamefont {X.}~\bibnamefont {Dai}}, \bibinfo {author} {\bibfnamefont {W.-Y.}\ \bibnamefont {Shan}}, \bibinfo {author} {\bibfnamefont {S.-Q.}\ \bibnamefont {Shen}}, \bibinfo {author} {\bibfnamefont {Q.}~\bibnamefont {Niu}}, \bibinfo {author} {\bibfnamefont {X.-L.}\ \bibnamefont {Qi}}, \bibinfo {author} {\bibfnamefont {S.-C.}\ \bibnamefont {Zhang}}, \bibinfo {author} {\bibfnamefont {X.-C.}\ \bibnamefont {Ma}},\ and\ \bibinfo {author} {\bibfnamefont {Q.-K.}\ \bibnamefont {Xue}},\
  }\href {https://doi.org/10.1038/nphys1689} {\bibfield  {journal} {\bibinfo  {journal} {Nat. Phys.}\ }\textbf {\bibinfo {volume} {6}},\ \bibinfo {pages} {584} (\bibinfo {year} {2010})}\BibitemShut {NoStop}%
\bibitem [{\citenamefont {Liu}\ \emph {et~al.}(2012)\citenamefont {Liu}, \citenamefont {Bian}, \citenamefont {Miller}, \citenamefont {Bissen},\ and\ \citenamefont {Chiang}}]{Liu2012}%
  \BibitemOpen
  \bibfield  {author} {\bibinfo {author} {\bibfnamefont {Y.}~\bibnamefont {Liu}}, \bibinfo {author} {\bibfnamefont {G.}~\bibnamefont {Bian}}, \bibinfo {author} {\bibfnamefont {T.}~\bibnamefont {Miller}}, \bibinfo {author} {\bibfnamefont {M.}~\bibnamefont {Bissen}},\ and\ \bibinfo {author} {\bibfnamefont {T.-C.}\ \bibnamefont {Chiang}},\ }\href {https://doi.org/10.1103/PhysRevB.85.195442} {\bibfield  {journal} {\bibinfo  {journal} {Phys. Rev. B}\ }\textbf {\bibinfo {volume} {85}},\ \bibinfo {pages} {195442} (\bibinfo {year} {2012})}\BibitemShut {NoStop}%
\bibitem [{\citenamefont {Jiang}\ \emph {et~al.}(2012)\citenamefont {Jiang}, \citenamefont {Wang}, \citenamefont {Chen}, \citenamefont {Li}, \citenamefont {Song}, \citenamefont {He}, \citenamefont {Wang}, \citenamefont {Chen}, \citenamefont {Ma},\ and\ \citenamefont {Xue}}]{Jiang2012}%
  \BibitemOpen
  \bibfield  {author} {\bibinfo {author} {\bibfnamefont {Y.}~\bibnamefont {Jiang}}, \bibinfo {author} {\bibfnamefont {Y.}~\bibnamefont {Wang}}, \bibinfo {author} {\bibfnamefont {M.}~\bibnamefont {Chen}}, \bibinfo {author} {\bibfnamefont {Z.}~\bibnamefont {Li}}, \bibinfo {author} {\bibfnamefont {C.}~\bibnamefont {Song}}, \bibinfo {author} {\bibfnamefont {K.}~\bibnamefont {He}}, \bibinfo {author} {\bibfnamefont {L.}~\bibnamefont {Wang}}, \bibinfo {author} {\bibfnamefont {X.}~\bibnamefont {Chen}}, \bibinfo {author} {\bibfnamefont {X.}~\bibnamefont {Ma}},\ and\ \bibinfo {author} {\bibfnamefont {Q.-K.}\ \bibnamefont {Xue}},\ }\href {https://doi.org/10.1103/PhysRevLett.108.016401} {\bibfield  {journal} {\bibinfo  {journal} {Phys. Rev. Lett.}\ }\textbf {\bibinfo {volume} {108}},\ \bibinfo {pages} {016401} (\bibinfo {year} {2012})}\BibitemShut {NoStop}%
\bibitem [{\citenamefont {Bernevig}\ \emph {et~al.}(2006)\citenamefont {Bernevig}, \citenamefont {Hughes},\ and\ \citenamefont {Zhang}}]{bhz}%
  \BibitemOpen
  \bibfield  {author} {\bibinfo {author} {\bibfnamefont {B.~A.}\ \bibnamefont {Bernevig}}, \bibinfo {author} {\bibfnamefont {T.~L.}\ \bibnamefont {Hughes}},\ and\ \bibinfo {author} {\bibfnamefont {S.-C.}\ \bibnamefont {Zhang}},\ }\href {https://doi.org/10.1126/science.1133734} {\bibfield  {journal} {\bibinfo  {journal} {Science}\ }\textbf {\bibinfo {volume} {314}},\ \bibinfo {pages} {1757} (\bibinfo {year} {2006})}\BibitemShut {NoStop}%
\bibitem [{\citenamefont {Groth}\ \emph {et~al.}(2014)\citenamefont {Groth}, \citenamefont {Wimmer}, \citenamefont {Akhmerov},\ and\ \citenamefont {Waintal}}]{Groth2014}%
  \BibitemOpen
  \bibfield  {author} {\bibinfo {author} {\bibfnamefont {C.~W.}\ \bibnamefont {Groth}}, \bibinfo {author} {\bibfnamefont {M.}~\bibnamefont {Wimmer}}, \bibinfo {author} {\bibfnamefont {A.~R.}\ \bibnamefont {Akhmerov}},\ and\ \bibinfo {author} {\bibfnamefont {X.}~\bibnamefont {Waintal}},\ }\href {https://doi.org/10.1088/1367-2630/16/6/063065} {\bibfield  {journal} {\bibinfo  {journal} {New J. Phys.}\ }\textbf {\bibinfo {volume} {16}},\ \bibinfo {pages} {063065} (\bibinfo {year} {2014})}\BibitemShut {NoStop}%
\bibitem [{\citenamefont {Nijholt}\ \emph {et~al.}(2019)\citenamefont {Nijholt}, \citenamefont {Weston}, \citenamefont {Hoofwijk},\ and\ \citenamefont {Akhmerov}}]{Nijholt2019}%
  \BibitemOpen
  \bibfield  {author} {\bibinfo {author} {\bibfnamefont {B.}~\bibnamefont {Nijholt}}, \bibinfo {author} {\bibfnamefont {J.}~\bibnamefont {Weston}}, \bibinfo {author} {\bibfnamefont {J.}~\bibnamefont {Hoofwijk}},\ and\ \bibinfo {author} {\bibfnamefont {A.}~\bibnamefont {Akhmerov}},\ }\href {https://doi.org/10.5281/zenodo.1182437} {\bibinfo {title} {\textit{Adaptive}: parallel active learning of mathematical functions}} (\bibinfo {year} {2019})\BibitemShut {NoStop}%
\bibitem [{\citenamefont {Zhang}\ \emph {et~al.}(2011)\citenamefont {Zhang}, \citenamefont {Chang}, \citenamefont {Zhang}, \citenamefont {Wen}, \citenamefont {Feng}, \citenamefont {Li}, \citenamefont {Liu}, \citenamefont {He}, \citenamefont {Wang}, \citenamefont {Chen}, \citenamefont {Xue}, \citenamefont {Ma},\ and\ \citenamefont {Wang}}]{Zhang2011}%
  \BibitemOpen
  \bibfield  {author} {\bibinfo {author} {\bibfnamefont {J.}~\bibnamefont {Zhang}}, \bibinfo {author} {\bibfnamefont {C.-Z.}\ \bibnamefont {Chang}}, \bibinfo {author} {\bibfnamefont {Z.}~\bibnamefont {Zhang}}, \bibinfo {author} {\bibfnamefont {J.}~\bibnamefont {Wen}}, \bibinfo {author} {\bibfnamefont {X.}~\bibnamefont {Feng}}, \bibinfo {author} {\bibfnamefont {K.}~\bibnamefont {Li}}, \bibinfo {author} {\bibfnamefont {M.}~\bibnamefont {Liu}}, \bibinfo {author} {\bibfnamefont {K.}~\bibnamefont {He}}, \bibinfo {author} {\bibfnamefont {L.}~\bibnamefont {Wang}}, \bibinfo {author} {\bibfnamefont {X.}~\bibnamefont {Chen}}, \bibinfo {author} {\bibfnamefont {Q.-K.}\ \bibnamefont {Xue}}, \bibinfo {author} {\bibfnamefont {X.}~\bibnamefont {Ma}},\ and\ \bibinfo {author} {\bibfnamefont {Y.}~\bibnamefont {Wang}},\ }\href {https://doi.org/10.1038/ncomms1588} {\bibfield  {journal} {\bibinfo  {journal} {Nat. Commun.}\ }\textbf {\bibinfo {volume} {2}},\ \bibinfo {pages} {574} (\bibinfo {year} {2011})}\BibitemShut {NoStop}%
\bibitem [{\citenamefont {Vidal}\ \emph {et~al.}(2013)\citenamefont {Vidal}, \citenamefont {Eddrief}, \citenamefont {Rache~Salles}, \citenamefont {Vobornik}, \citenamefont {Velez-Fort}, \citenamefont {Panaccione},\ and\ \citenamefont {Marangolo}}]{Vidal_2013}%
  \BibitemOpen
  \bibfield  {author} {\bibinfo {author} {\bibfnamefont {F.}~\bibnamefont {Vidal}}, \bibinfo {author} {\bibfnamefont {M.}~\bibnamefont {Eddrief}}, \bibinfo {author} {\bibfnamefont {B.}~\bibnamefont {Rache~Salles}}, \bibinfo {author} {\bibfnamefont {I.}~\bibnamefont {Vobornik}}, \bibinfo {author} {\bibfnamefont {E.}~\bibnamefont {Velez-Fort}}, \bibinfo {author} {\bibfnamefont {G.}~\bibnamefont {Panaccione}},\ and\ \bibinfo {author} {\bibfnamefont {M.}~\bibnamefont {Marangolo}},\ }\href {https://doi.org/10.1103/PhysRevB.88.241410} {\bibfield  {journal} {\bibinfo  {journal} {Phys. Rev. B}\ }\textbf {\bibinfo {volume} {88}},\ \bibinfo {pages} {241410} (\bibinfo {year} {2013})}\BibitemShut {NoStop}%
\bibitem [{\citenamefont {Landolt}\ \emph {et~al.}(2014)\citenamefont {Landolt}, \citenamefont {Schreyeck}, \citenamefont {Eremeev}, \citenamefont {Slomski}, \citenamefont {Muff}, \citenamefont {Osterwalder}, \citenamefont {Chulkov}, \citenamefont {Gould}, \citenamefont {Karczewski}, \citenamefont {Brunner}, \citenamefont {Buhmann}, \citenamefont {Molenkamp},\ and\ \citenamefont {Dil}}]{Landolt_2014}%
  \BibitemOpen
  \bibfield  {author} {\bibinfo {author} {\bibfnamefont {G.}~\bibnamefont {Landolt}}, \bibinfo {author} {\bibfnamefont {S.}~\bibnamefont {Schreyeck}}, \bibinfo {author} {\bibfnamefont {S.~V.}\ \bibnamefont {Eremeev}}, \bibinfo {author} {\bibfnamefont {B.}~\bibnamefont {Slomski}}, \bibinfo {author} {\bibfnamefont {S.}~\bibnamefont {Muff}}, \bibinfo {author} {\bibfnamefont {J.}~\bibnamefont {Osterwalder}}, \bibinfo {author} {\bibfnamefont {E.~V.}\ \bibnamefont {Chulkov}}, \bibinfo {author} {\bibfnamefont {C.}~\bibnamefont {Gould}}, \bibinfo {author} {\bibfnamefont {G.}~\bibnamefont {Karczewski}}, \bibinfo {author} {\bibfnamefont {K.}~\bibnamefont {Brunner}}, \bibinfo {author} {\bibfnamefont {H.}~\bibnamefont {Buhmann}}, \bibinfo {author} {\bibfnamefont {L.~W.}\ \bibnamefont {Molenkamp}},\ and\ \bibinfo {author} {\bibfnamefont {J.~H.}\ \bibnamefont {Dil}},\ }\href {https://doi.org/10.1103/PhysRevLett.112.057601} {\bibfield  {journal} {\bibinfo  {journal} {Phys. Rev. Lett.}\ }\textbf {\bibinfo {volume} {112}},\
  \bibinfo {pages} {057601} (\bibinfo {year} {2014})}\BibitemShut {NoStop}%
\bibitem [{\citenamefont {Neupane}\ \emph {et~al.}(2014)\citenamefont {Neupane}, \citenamefont {Richardella}, \citenamefont {S{\'{a}}nchez-Barriga}, \citenamefont {Xu}, \citenamefont {Alidoust}, \citenamefont {Belopolski}, \citenamefont {Liu}, \citenamefont {Bian}, \citenamefont {Zhang}, \citenamefont {Marchenko}, \citenamefont {Varykhalov}, \citenamefont {Rader}, \citenamefont {Leandersson}, \citenamefont {Balasubramanian}, \citenamefont {Chang}, \citenamefont {Jeng}, \citenamefont {Basak}, \citenamefont {Lin}, \citenamefont {Bansil}, \citenamefont {Samarth},\ and\ \citenamefont {Hasan}}]{Neupane_2014}%
  \BibitemOpen
  \bibfield  {author} {\bibinfo {author} {\bibfnamefont {M.}~\bibnamefont {Neupane}}, \bibinfo {author} {\bibfnamefont {A.}~\bibnamefont {Richardella}}, \bibinfo {author} {\bibfnamefont {J.}~\bibnamefont {S{\'{a}}nchez-Barriga}}, \bibinfo {author} {\bibfnamefont {S.}~\bibnamefont {Xu}}, \bibinfo {author} {\bibfnamefont {N.}~\bibnamefont {Alidoust}}, \bibinfo {author} {\bibfnamefont {I.}~\bibnamefont {Belopolski}}, \bibinfo {author} {\bibfnamefont {C.}~\bibnamefont {Liu}}, \bibinfo {author} {\bibfnamefont {G.}~\bibnamefont {Bian}}, \bibinfo {author} {\bibfnamefont {D.}~\bibnamefont {Zhang}}, \bibinfo {author} {\bibfnamefont {D.}~\bibnamefont {Marchenko}}, \bibinfo {author} {\bibfnamefont {A.}~\bibnamefont {Varykhalov}}, \bibinfo {author} {\bibfnamefont {O.}~\bibnamefont {Rader}}, \bibinfo {author} {\bibfnamefont {M.}~\bibnamefont {Leandersson}}, \bibinfo {author} {\bibfnamefont {T.}~\bibnamefont {Balasubramanian}}, \bibinfo {author} {\bibfnamefont {T.-R.}\ \bibnamefont {Chang}}, \bibinfo {author}
  {\bibfnamefont {H.-T.}\ \bibnamefont {Jeng}}, \bibinfo {author} {\bibfnamefont {S.}~\bibnamefont {Basak}}, \bibinfo {author} {\bibfnamefont {H.}~\bibnamefont {Lin}}, \bibinfo {author} {\bibfnamefont {A.}~\bibnamefont {Bansil}}, \bibinfo {author} {\bibfnamefont {N.}~\bibnamefont {Samarth}},\ and\ \bibinfo {author} {\bibfnamefont {M.~Z.}\ \bibnamefont {Hasan}},\ }\href {https://doi.org/10.1038/ncomms4841} {\bibfield  {journal} {\bibinfo  {journal} {Nat. Commun.}\ }\textbf {\bibinfo {volume} {5}},\ \bibinfo {pages} {3841} (\bibinfo {year} {2014})}\BibitemShut {NoStop}%
\bibitem [{\citenamefont {Kong}\ \emph {et~al.}(2010)\citenamefont {Kong}, \citenamefont {Randel}, \citenamefont {Peng}, \citenamefont {Cha}, \citenamefont {Meister}, \citenamefont {Lai}, \citenamefont {Chen}, \citenamefont {Shen}, \citenamefont {Manoharan},\ and\ \citenamefont {Cui}}]{kong2010}%
  \BibitemOpen
  \bibfield  {author} {\bibinfo {author} {\bibfnamefont {D.}~\bibnamefont {Kong}}, \bibinfo {author} {\bibfnamefont {J.~C.}\ \bibnamefont {Randel}}, \bibinfo {author} {\bibfnamefont {H.}~\bibnamefont {Peng}}, \bibinfo {author} {\bibfnamefont {J.~J.}\ \bibnamefont {Cha}}, \bibinfo {author} {\bibfnamefont {S.}~\bibnamefont {Meister}}, \bibinfo {author} {\bibfnamefont {K.}~\bibnamefont {Lai}}, \bibinfo {author} {\bibfnamefont {Y.}~\bibnamefont {Chen}}, \bibinfo {author} {\bibfnamefont {Z.-X.}\ \bibnamefont {Shen}}, \bibinfo {author} {\bibfnamefont {H.~C.}\ \bibnamefont {Manoharan}},\ and\ \bibinfo {author} {\bibfnamefont {Y.}~\bibnamefont {Cui}},\ }\href {https://doi.org/10.1021/nl903663a} {\bibfield  {journal} {\bibinfo  {journal} {Nano Lett.}\ }\textbf {\bibinfo {volume} {10}},\ \bibinfo {pages} {329} (\bibinfo {year} {2010})}\BibitemShut {NoStop}%
\bibitem [{\citenamefont {Xiu}\ \emph {et~al.}(2011)\citenamefont {Xiu}, \citenamefont {He}, \citenamefont {Wang}, \citenamefont {Cheng}, \citenamefont {Chang}, \citenamefont {Lang}, \citenamefont {Huang}, \citenamefont {Kou}, \citenamefont {Zhou}, \citenamefont {Jiang}, \citenamefont {Chen}, \citenamefont {Zou}, \citenamefont {Shailos},\ and\ \citenamefont {Wang}}]{Xiu_2011}%
  \BibitemOpen
  \bibfield  {author} {\bibinfo {author} {\bibfnamefont {F.}~\bibnamefont {Xiu}}, \bibinfo {author} {\bibfnamefont {L.}~\bibnamefont {He}}, \bibinfo {author} {\bibfnamefont {Y.}~\bibnamefont {Wang}}, \bibinfo {author} {\bibfnamefont {L.}~\bibnamefont {Cheng}}, \bibinfo {author} {\bibfnamefont {L.-T.}\ \bibnamefont {Chang}}, \bibinfo {author} {\bibfnamefont {M.}~\bibnamefont {Lang}}, \bibinfo {author} {\bibfnamefont {G.}~\bibnamefont {Huang}}, \bibinfo {author} {\bibfnamefont {X.}~\bibnamefont {Kou}}, \bibinfo {author} {\bibfnamefont {Y.}~\bibnamefont {Zhou}}, \bibinfo {author} {\bibfnamefont {X.}~\bibnamefont {Jiang}}, \bibinfo {author} {\bibfnamefont {Z.}~\bibnamefont {Chen}}, \bibinfo {author} {\bibfnamefont {J.}~\bibnamefont {Zou}}, \bibinfo {author} {\bibfnamefont {A.}~\bibnamefont {Shailos}},\ and\ \bibinfo {author} {\bibfnamefont {K.~L.}\ \bibnamefont {Wang}},\ }\href {https://doi.org/10.1038/nnano.2011.19} {\bibfield  {journal} {\bibinfo  {journal} {Nat. Nanotechnol.}\ }\textbf {\bibinfo {volume}
  {6}},\ \bibinfo {pages} {216–221} (\bibinfo {year} {2011})}\BibitemShut {NoStop}%
\bibitem [{\citenamefont {Zhang}\ and\ \citenamefont {Vishwanath}(2010)}]{Zhang2010B}%
  \BibitemOpen
  \bibfield  {author} {\bibinfo {author} {\bibfnamefont {Y.}~\bibnamefont {Zhang}}\ and\ \bibinfo {author} {\bibfnamefont {A.}~\bibnamefont {Vishwanath}},\ }\href {https://doi.org/10.1103/PhysRevLett.105.206601} {\bibfield  {journal} {\bibinfo  {journal} {Phys. Rev. Lett.}\ }\textbf {\bibinfo {volume} {105}},\ \bibinfo {pages} {206601} (\bibinfo {year} {2010})}\BibitemShut {NoStop}%
\bibitem [{\citenamefont {F\"orster}(2016)}]{Forster_phd}%
  \BibitemOpen
  \bibfield  {author} {\bibinfo {author} {\bibfnamefont {T.}~\bibnamefont {F\"orster}},\ }\emph {\bibinfo {title} {Topological insulators ${Bi}_{2}{Se}_{3}$, ${Bi}_{2}{Te}_{3}$, and ${Sb}_{2}{Te}_{3}$: Electronic and topological properties of surfaces and thin films from DFT and GW calculations}},\ \href@noop {} {\bibinfo {type} {Phd thesis}},\ \bibinfo  {school} {University of Münster} (\bibinfo {year} {2016})\BibitemShut {NoStop}%
\bibitem [{\citenamefont {Lu}\ \emph {et~al.}(2013)\citenamefont {Lu}, \citenamefont {Zhao},\ and\ \citenamefont {Shen}}]{Lu_2013}%
  \BibitemOpen
  \bibfield  {author} {\bibinfo {author} {\bibfnamefont {H.-Z.}\ \bibnamefont {Lu}}, \bibinfo {author} {\bibfnamefont {A.}~\bibnamefont {Zhao}},\ and\ \bibinfo {author} {\bibfnamefont {S.-Q.}\ \bibnamefont {Shen}},\ }\href {https://doi.org/10.1103/PhysRevLett.111.146802} {\bibfield  {journal} {\bibinfo  {journal} {Phys. Rev. Lett.}\ }\textbf {\bibinfo {volume} {111}},\ \bibinfo {pages} {146802} (\bibinfo {year} {2013})}\BibitemShut {NoStop}%
\bibitem [{\citenamefont {Rosenbach}\ \emph {et~al.}(2022)\citenamefont {Rosenbach}, \citenamefont {Moors}, \citenamefont {Jalil}, \citenamefont {Kölzer}, \citenamefont {Zimmermann}, \citenamefont {Schubert}, \citenamefont {Karimzadah}, \citenamefont {Mussler}, \citenamefont {Schüffelgen}, \citenamefont {Grützmacher}, \citenamefont {Lüth},\ and\ \citenamefont {Schäpers}}]{rosenbach_2022}%
  \BibitemOpen
  \bibfield  {author} {\bibinfo {author} {\bibfnamefont {D.}~\bibnamefont {Rosenbach}}, \bibinfo {author} {\bibfnamefont {K.}~\bibnamefont {Moors}}, \bibinfo {author} {\bibfnamefont {A.~R.}\ \bibnamefont {Jalil}}, \bibinfo {author} {\bibfnamefont {J.}~\bibnamefont {Kölzer}}, \bibinfo {author} {\bibfnamefont {E.}~\bibnamefont {Zimmermann}}, \bibinfo {author} {\bibfnamefont {J.}~\bibnamefont {Schubert}}, \bibinfo {author} {\bibfnamefont {S.}~\bibnamefont {Karimzadah}}, \bibinfo {author} {\bibfnamefont {G.}~\bibnamefont {Mussler}}, \bibinfo {author} {\bibfnamefont {P.}~\bibnamefont {Schüffelgen}}, \bibinfo {author} {\bibfnamefont {D.}~\bibnamefont {Grützmacher}}, \bibinfo {author} {\bibfnamefont {H.}~\bibnamefont {Lüth}},\ and\ \bibinfo {author} {\bibfnamefont {T.}~\bibnamefont {Schäpers}},\ }\href {https://doi.org/10.21468/SciPostPhysCore.5.1.017} {\bibfield  {journal} {\bibinfo  {journal} {SciPost Phys. Core}\ }\textbf {\bibinfo {volume} {5}},\ \bibinfo {pages} {017} (\bibinfo {year} {2022})}\BibitemShut
  {NoStop}%
\bibitem [{\citenamefont {Sitthison}\ and\ \citenamefont {Stanescu}(2014)}]{sitthison2014}%
  \BibitemOpen
  \bibfield  {author} {\bibinfo {author} {\bibfnamefont {P.}~\bibnamefont {Sitthison}}\ and\ \bibinfo {author} {\bibfnamefont {T.~D.}\ \bibnamefont {Stanescu}},\ }\href {https://doi.org/10.1103/PhysRevB.90.035313} {\bibfield  {journal} {\bibinfo  {journal} {Phys. Rev. B}\ }\textbf {\bibinfo {volume} {90}},\ \bibinfo {pages} {035313} (\bibinfo {year} {2014})}\BibitemShut {NoStop}%
\bibitem [{\citenamefont {Heffels}\ \emph {et~al.}(2023)\citenamefont {Heffels}, \citenamefont {Burke}, \citenamefont {Connolly}, \citenamefont {Schüffelgen}, \citenamefont {Grützmacher},\ and\ \citenamefont {Moors}}]{Heffels2023}%
  \BibitemOpen
  \bibfield  {author} {\bibinfo {author} {\bibfnamefont {D.}~\bibnamefont {Heffels}}, \bibinfo {author} {\bibfnamefont {D.}~\bibnamefont {Burke}}, \bibinfo {author} {\bibfnamefont {M.~R.}\ \bibnamefont {Connolly}}, \bibinfo {author} {\bibfnamefont {P.}~\bibnamefont {Schüffelgen}}, \bibinfo {author} {\bibfnamefont {D.}~\bibnamefont {Grützmacher}},\ and\ \bibinfo {author} {\bibfnamefont {K.}~\bibnamefont {Moors}},\ }\href {https://doi.org/10.3390/nano13040723} {\bibfield  {journal} {\bibinfo  {journal} {Nanomaterials}\ }\textbf {\bibinfo {volume} {13}},\ \bibinfo {pages} {723} (\bibinfo {year} {2023})}\BibitemShut {NoStop}%
\bibitem [{\citenamefont {Fu}\ and\ \citenamefont {Kane}(2008)}]{Fu2008}%
  \BibitemOpen
  \bibfield  {author} {\bibinfo {author} {\bibfnamefont {L.}~\bibnamefont {Fu}}\ and\ \bibinfo {author} {\bibfnamefont {C.~L.}\ \bibnamefont {Kane}},\ }\href {https://doi.org/10.1103/PhysRevLett.100.096407} {\bibfield  {journal} {\bibinfo  {journal} {Phys. Rev. Lett.}\ }\textbf {\bibinfo {volume} {100}},\ \bibinfo {pages} {096407} (\bibinfo {year} {2008})}\BibitemShut {NoStop}%
\bibitem [{\citenamefont {Stanescu}\ \emph {et~al.}(2010)\citenamefont {Stanescu}, \citenamefont {Sau}, \citenamefont {Lutchyn},\ and\ \citenamefont {Das~Sarma}}]{Stanescu2010}%
  \BibitemOpen
  \bibfield  {author} {\bibinfo {author} {\bibfnamefont {T.~D.}\ \bibnamefont {Stanescu}}, \bibinfo {author} {\bibfnamefont {J.~D.}\ \bibnamefont {Sau}}, \bibinfo {author} {\bibfnamefont {R.~M.}\ \bibnamefont {Lutchyn}},\ and\ \bibinfo {author} {\bibfnamefont {S.}~\bibnamefont {Das~Sarma}},\ }\href {https://doi.org/10.1103/PhysRevB.81.241310} {\bibfield  {journal} {\bibinfo  {journal} {Phys. Rev. B}\ }\textbf {\bibinfo {volume} {81}},\ \bibinfo {pages} {241310} (\bibinfo {year} {2010})}\BibitemShut {NoStop}%
\bibitem [{\citenamefont {Potter}\ and\ \citenamefont {Lee}(2011)}]{potter2011}%
  \BibitemOpen
  \bibfield  {author} {\bibinfo {author} {\bibfnamefont {A.~C.}\ \bibnamefont {Potter}}\ and\ \bibinfo {author} {\bibfnamefont {P.~A.}\ \bibnamefont {Lee}},\ }\href {https://doi.org/10.1103/PhysRevB.83.184520} {\bibfield  {journal} {\bibinfo  {journal} {Phys. Rev. B}\ }\textbf {\bibinfo {volume} {83}},\ \bibinfo {pages} {184520} (\bibinfo {year} {2011})}\BibitemShut {NoStop}%
\bibitem [{\citenamefont {Cook}\ and\ \citenamefont {Franz}(2011)}]{cook2011}%
  \BibitemOpen
  \bibfield  {author} {\bibinfo {author} {\bibfnamefont {A.}~\bibnamefont {Cook}}\ and\ \bibinfo {author} {\bibfnamefont {M.}~\bibnamefont {Franz}},\ }\href {https://doi.org/10.1103/PhysRevB.84.201105} {\bibfield  {journal} {\bibinfo  {journal} {Phys. Rev. B}\ }\textbf {\bibinfo {volume} {84}},\ \bibinfo {pages} {201105} (\bibinfo {year} {2011})}\BibitemShut {NoStop}%
\bibitem [{\citenamefont {Cook}\ \emph {et~al.}(2012)\citenamefont {Cook}, \citenamefont {Vazifeh},\ and\ \citenamefont {Franz}}]{cook2012}%
  \BibitemOpen
  \bibfield  {author} {\bibinfo {author} {\bibfnamefont {A.~M.}\ \bibnamefont {Cook}}, \bibinfo {author} {\bibfnamefont {M.~M.}\ \bibnamefont {Vazifeh}},\ and\ \bibinfo {author} {\bibfnamefont {M.}~\bibnamefont {Franz}},\ }\href {https://doi.org/10.1103/PhysRevB.86.155431} {\bibfield  {journal} {\bibinfo  {journal} {Phys. Rev. B}\ }\textbf {\bibinfo {volume} {86}},\ \bibinfo {pages} {155431} (\bibinfo {year} {2012})}\BibitemShut {NoStop}%
\bibitem [{\citenamefont {Chiu}\ \emph {et~al.}(2016)\citenamefont {Chiu}, \citenamefont {Cole},\ and\ \citenamefont {Das~Sarma}}]{chiu2016}%
  \BibitemOpen
  \bibfield  {author} {\bibinfo {author} {\bibfnamefont {C.-K.}\ \bibnamefont {Chiu}}, \bibinfo {author} {\bibfnamefont {W.~S.}\ \bibnamefont {Cole}},\ and\ \bibinfo {author} {\bibfnamefont {S.}~\bibnamefont {Das~Sarma}},\ }\href {https://doi.org/10.1103/PhysRevB.94.125304} {\bibfield  {journal} {\bibinfo  {journal} {Phys. Rev. B}\ }\textbf {\bibinfo {volume} {94}},\ \bibinfo {pages} {125304} (\bibinfo {year} {2016})}\BibitemShut {NoStop}%
\bibitem [{\citenamefont {Bai}\ \emph {et~al.}(2022)\citenamefont {Bai}, \citenamefont {Wei}, \citenamefont {Feng}, \citenamefont {Luysberg}, \citenamefont {Bliesener}, \citenamefont {Lippertz}, \citenamefont {Uday}, \citenamefont {Taskin}, \citenamefont {Mayer},\ and\ \citenamefont {Ando}}]{Bai_2022}%
  \BibitemOpen
  \bibfield  {author} {\bibinfo {author} {\bibfnamefont {M.}~\bibnamefont {Bai}}, \bibinfo {author} {\bibfnamefont {X.-K.}\ \bibnamefont {Wei}}, \bibinfo {author} {\bibfnamefont {J.}~\bibnamefont {Feng}}, \bibinfo {author} {\bibfnamefont {M.}~\bibnamefont {Luysberg}}, \bibinfo {author} {\bibfnamefont {A.}~\bibnamefont {Bliesener}}, \bibinfo {author} {\bibfnamefont {G.}~\bibnamefont {Lippertz}}, \bibinfo {author} {\bibfnamefont {A.}~\bibnamefont {Uday}}, \bibinfo {author} {\bibfnamefont {A.~A.}\ \bibnamefont {Taskin}}, \bibinfo {author} {\bibfnamefont {J.}~\bibnamefont {Mayer}},\ and\ \bibinfo {author} {\bibfnamefont {Y.}~\bibnamefont {Ando}},\ }\href {https://doi.org/10.1038/s43246-022-00242-6} {\bibfield  {journal} {\bibinfo  {journal} {Commun. Mater.}\ }\textbf {\bibinfo {volume} {3}},\ \bibinfo {pages} {20} (\bibinfo {year} {2022})}\BibitemShut {NoStop}%
\bibitem [{\citenamefont {Burke}\ \emph {et~al.}(2024)\citenamefont {Burke}, \citenamefont {Heffels}, \citenamefont {Moors}, \citenamefont {Sch\"uffelgen}, \citenamefont {Gr\"utzmacher},\ and\ \citenamefont {Connolly}}]{Burke2024}%
  \BibitemOpen
  \bibfield  {author} {\bibinfo {author} {\bibfnamefont {D.}~\bibnamefont {Burke}}, \bibinfo {author} {\bibfnamefont {D.}~\bibnamefont {Heffels}}, \bibinfo {author} {\bibfnamefont {K.}~\bibnamefont {Moors}}, \bibinfo {author} {\bibfnamefont {P.}~\bibnamefont {Sch\"uffelgen}}, \bibinfo {author} {\bibfnamefont {D.}~\bibnamefont {Gr\"utzmacher}},\ and\ \bibinfo {author} {\bibfnamefont {M.~R.}\ \bibnamefont {Connolly}},\ }\href {https://doi.org/10.1103/PhysRevB.109.045138} {\bibfield  {journal} {\bibinfo  {journal} {Phys. Rev. B}\ }\textbf {\bibinfo {volume} {109}},\ \bibinfo {pages} {045138} (\bibinfo {year} {2024})}\BibitemShut {NoStop}%
\bibitem [{\citenamefont {Ebert}\ \emph {et~al.}(2011)\citenamefont {Ebert}, \citenamefont {K{\"o}dderitzsch},\ and\ \citenamefont {Min{\'a}r}}]{Ebert2011}%
  \BibitemOpen
  \bibfield  {author} {\bibinfo {author} {\bibfnamefont {H.}~\bibnamefont {Ebert}}, \bibinfo {author} {\bibfnamefont {D.}~\bibnamefont {K{\"o}dderitzsch}},\ and\ \bibinfo {author} {\bibfnamefont {J.}~\bibnamefont {Min{\'a}r}},\ }\href {https://doi.org/10.1088/0034-4885/74/9/096501} {\bibfield  {journal} {\bibinfo  {journal} {Rep. Prog. Phys.}\ }\textbf {\bibinfo {volume} {74}},\ \bibinfo {pages} {096501} (\bibinfo {year} {2011})}\BibitemShut {NoStop}%
\bibitem [{\citenamefont {Rüßmann}\ \emph {et~al.}(2022)\citenamefont {Rüßmann}, \citenamefont {Silva}, \citenamefont {Bauer}, \citenamefont {Baumeister}, \citenamefont {Bornemann}, \citenamefont {Bouaziz}, \citenamefont {Brinker}, \citenamefont {Chico}, \citenamefont {Dederichs}, \citenamefont {Drittler}, \citenamefont {Santos}, \citenamefont {dos Santos~Dias}, \citenamefont {Essing}, \citenamefont {Géranton}, \citenamefont {Klepetsanis}, \citenamefont {Kosma}, \citenamefont {Long}, \citenamefont {Lounis}, \citenamefont {Mavropoulos}, \citenamefont {Tapia}, \citenamefont {Oran}, \citenamefont {Papanikolaou}, \citenamefont {Rabel}, \citenamefont {Schweflinghaus}, \citenamefont {Stefanou}, \citenamefont {Thiess}, \citenamefont {Zeller}, \citenamefont {Zimmermann},\ and\ \citenamefont {Blügel}}]{jukkr2022}%
  \BibitemOpen
  \bibfield  {author} {\bibinfo {author} {\bibfnamefont {P.}~\bibnamefont {Rüßmann}}, \bibinfo {author} {\bibfnamefont {D.~A.}\ \bibnamefont {Silva}}, \bibinfo {author} {\bibfnamefont {D.}~\bibnamefont {Bauer}}, \bibinfo {author} {\bibfnamefont {P.}~\bibnamefont {Baumeister}}, \bibinfo {author} {\bibfnamefont {M.}~\bibnamefont {Bornemann}}, \bibinfo {author} {\bibfnamefont {J.}~\bibnamefont {Bouaziz}}, \bibinfo {author} {\bibfnamefont {S.}~\bibnamefont {Brinker}}, \bibinfo {author} {\bibfnamefont {J.}~\bibnamefont {Chico}}, \bibinfo {author} {\bibfnamefont {P.}~\bibnamefont {Dederichs}}, \bibinfo {author} {\bibfnamefont {B.}~\bibnamefont {Drittler}}, \bibinfo {author} {\bibfnamefont {F.~D.}\ \bibnamefont {Santos}}, \bibinfo {author} {\bibfnamefont {M.}~\bibnamefont {dos Santos~Dias}}, \bibinfo {author} {\bibfnamefont {N.}~\bibnamefont {Essing}}, \bibinfo {author} {\bibfnamefont {G.}~\bibnamefont {Géranton}}, \bibinfo {author} {\bibfnamefont {I.}~\bibnamefont {Klepetsanis}}, \bibinfo {author} {\bibfnamefont
  {A.}~\bibnamefont {Kosma}}, \bibinfo {author} {\bibfnamefont {N.}~\bibnamefont {Long}}, \bibinfo {author} {\bibfnamefont {S.}~\bibnamefont {Lounis}}, \bibinfo {author} {\bibfnamefont {P.}~\bibnamefont {Mavropoulos}}, \bibinfo {author} {\bibfnamefont {E.~M.}\ \bibnamefont {Tapia}}, \bibinfo {author} {\bibfnamefont {C.}~\bibnamefont {Oran}}, \bibinfo {author} {\bibfnamefont {N.}~\bibnamefont {Papanikolaou}}, \bibinfo {author} {\bibfnamefont {E.}~\bibnamefont {Rabel}}, \bibinfo {author} {\bibfnamefont {B.}~\bibnamefont {Schweflinghaus}}, \bibinfo {author} {\bibfnamefont {N.}~\bibnamefont {Stefanou}}, \bibinfo {author} {\bibfnamefont {A.}~\bibnamefont {Thiess}}, \bibinfo {author} {\bibfnamefont {R.}~\bibnamefont {Zeller}}, \bibinfo {author} {\bibfnamefont {B.}~\bibnamefont {Zimmermann}},\ and\ \bibinfo {author} {\bibfnamefont {S.}~\bibnamefont {Blügel}},\ }\bibfield  {journal} {\bibinfo  {journal} {Zenodo}\ }\href {https://doi.org/10.5281/zenodo.7284738} {10.5281/zenodo.7284738} (\bibinfo {year}
  {2022})\BibitemShut {NoStop}%
\bibitem [{\citenamefont {Nakajima}(1963)}]{Bi2Te3}%
  \BibitemOpen
  \bibfield  {author} {\bibinfo {author} {\bibfnamefont {S.}~\bibnamefont {Nakajima}},\ }\href {https://doi.org/10.1016/0022-3697(63)90207-5} {\bibfield  {journal} {\bibinfo  {journal} {J. Phys. Chem. Solids}\ ,\ \bibinfo {pages} {479}} (\bibinfo {year} {1963})}\BibitemShut {NoStop}%
\bibitem [{\citenamefont {Ullner}(1968)}]{Sb2Te3}%
  \BibitemOpen
  \bibfield  {author} {\bibinfo {author} {\bibfnamefont {H.-A.}\ \bibnamefont {Ullner}},\ }\href {https://doi.org/10.1002/andp.19684760106} {\bibfield  {journal} {\bibinfo  {journal} {Annalen der Physik}\ }\textbf {\bibinfo {volume} {21}},\ \bibinfo {pages} {45} (\bibinfo {year} {1968})}\BibitemShut {NoStop}%
\bibitem [{\citenamefont {Wyckoff}(1964)}]{Bi2Se3}%
  \BibitemOpen
  \bibfield  {author} {\bibinfo {author} {\bibfnamefont {R.~W.~G.}\ \bibnamefont {Wyckoff}},\ }\href@noop {} {\emph {\bibinfo {title} {Crystal Structures (2 ed.)}}}\ (\bibinfo  {publisher} {J. Wiley and Sons},\ \bibinfo {year} {1964})\BibitemShut {NoStop}%
\bibitem [{\citenamefont {Vosko}\ \emph {et~al.}(1980)\citenamefont {Vosko}, \citenamefont {Wilk},\ and\ \citenamefont {Nusair}}]{Vosko1980}%
  \BibitemOpen
  \bibfield  {author} {\bibinfo {author} {\bibfnamefont {S.~H.}\ \bibnamefont {Vosko}}, \bibinfo {author} {\bibfnamefont {L.}~\bibnamefont {Wilk}},\ and\ \bibinfo {author} {\bibfnamefont {M.}~\bibnamefont {Nusair}},\ }\href {https://doi.org/10.1139/p80-159} {\bibfield  {journal} {\bibinfo  {journal} {Can. J. Phys.}\ }\textbf {\bibinfo {volume} {58}},\ \bibinfo {pages} {1200} (\bibinfo {year} {1980})}\BibitemShut {NoStop}%
\bibitem [{\citenamefont {Perdew}\ \emph {et~al.}(1996)\citenamefont {Perdew}, \citenamefont {Burke},\ and\ \citenamefont {Ernzerhof}}]{PBE}%
  \BibitemOpen
  \bibfield  {author} {\bibinfo {author} {\bibfnamefont {J.~P.}\ \bibnamefont {Perdew}}, \bibinfo {author} {\bibfnamefont {K.}~\bibnamefont {Burke}},\ and\ \bibinfo {author} {\bibfnamefont {M.}~\bibnamefont {Ernzerhof}},\ }\href {https://doi.org/10.1103/PhysRevLett.77.3865} {\bibfield  {journal} {\bibinfo  {journal} {Phys. Rev. Lett.}\ }\textbf {\bibinfo {volume} {77}},\ \bibinfo {pages} {3865} (\bibinfo {year} {1996})}\BibitemShut {NoStop}%
\bibitem [{\citenamefont {Zeller}(2004)}]{Zeller2004}%
  \BibitemOpen
  \bibfield  {author} {\bibinfo {author} {\bibfnamefont {R.}~\bibnamefont {Zeller}},\ }\href {https://doi.org/10.1088/0953-8984/16/36/011} {\bibfield  {journal} {\bibinfo  {journal} {J. Phys.: Condens. Matter}\ }\textbf {\bibinfo {volume} {16}},\ \bibinfo {pages} {6453} (\bibinfo {year} {2004})}\BibitemShut {NoStop}%
\bibitem [{\citenamefont {Stefanou}\ \emph {et~al.}(1990)\citenamefont {Stefanou}, \citenamefont {Akai},\ and\ \citenamefont {Zeller}}]{Stefanou1990}%
  \BibitemOpen
  \bibfield  {author} {\bibinfo {author} {\bibfnamefont {N.}~\bibnamefont {Stefanou}}, \bibinfo {author} {\bibfnamefont {H.}~\bibnamefont {Akai}},\ and\ \bibinfo {author} {\bibfnamefont {R.}~\bibnamefont {Zeller}},\ }\href {https://doi.org/10.1016/0010-4655(90)90009-P} {\bibfield  {journal} {\bibinfo  {journal} {Comput. Phys. Commun.}\ }\textbf {\bibinfo {volume} {60}},\ \bibinfo {pages} {231} (\bibinfo {year} {1990})}\BibitemShut {NoStop}%
\bibitem [{\citenamefont {Stefanou}\ and\ \citenamefont {Zeller}(1991)}]{Stefanou1991}%
  \BibitemOpen
  \bibfield  {author} {\bibinfo {author} {\bibfnamefont {N.}~\bibnamefont {Stefanou}}\ and\ \bibinfo {author} {\bibfnamefont {R.}~\bibnamefont {Zeller}},\ }\href {https://doi.org/10.1088/0953-8984/3/39/006} {\bibfield  {journal} {\bibinfo  {journal} {J. Phys.: Cond. Matter}\ }\textbf {\bibinfo {volume} {3}},\ \bibinfo {pages} {7599} (\bibinfo {year} {1991})}\BibitemShut {NoStop}%
\bibitem [{\citenamefont {Rüßmann}\ \emph {et~al.}(2021)\citenamefont {Rüßmann}, \citenamefont {Bertoldo},\ and\ \citenamefont {Blügel}}]{aiida-kkr-paper}%
  \BibitemOpen
  \bibfield  {author} {\bibinfo {author} {\bibfnamefont {P.}~\bibnamefont {Rüßmann}}, \bibinfo {author} {\bibfnamefont {F.}~\bibnamefont {Bertoldo}},\ and\ \bibinfo {author} {\bibfnamefont {S.}~\bibnamefont {Blügel}},\ }\href {https://doi.org/10.1038/s41524-020-00482-5} {\bibfield  {journal} {\bibinfo  {journal} {npj Comput. Mater.}\ }\textbf {\bibinfo {volume} {7}},\ \bibinfo {pages} {13} (\bibinfo {year} {2021})}\BibitemShut {NoStop}%
\bibitem [{\citenamefont {Huber}\ \emph {et~al.}(2020)\citenamefont {Huber}, \citenamefont {Zoupanos}, \citenamefont {Uhrin}, \citenamefont {Talirz}, \citenamefont {Kahle}, \citenamefont {Häuselmann}, \citenamefont {Gresch}, \citenamefont {Müller}, \citenamefont {Yakutovich}, \citenamefont {Andersen}, \citenamefont {Ramirez}, \citenamefont {Adorf}, \citenamefont {Gargiulo}, \citenamefont {Kumbhar}, \citenamefont {Passaro}, \citenamefont {Johnston}, \citenamefont {Merkys}, \citenamefont {Cepellotti}, \citenamefont {Mounet}, \citenamefont {Marzari}, \citenamefont {Kozinsky},\ and\ \citenamefont {Pizzi}}]{aiida}%
  \BibitemOpen
  \bibfield  {author} {\bibinfo {author} {\bibfnamefont {S.~P.}\ \bibnamefont {Huber}}, \bibinfo {author} {\bibfnamefont {S.}~\bibnamefont {Zoupanos}}, \bibinfo {author} {\bibfnamefont {M.}~\bibnamefont {Uhrin}}, \bibinfo {author} {\bibfnamefont {L.}~\bibnamefont {Talirz}}, \bibinfo {author} {\bibfnamefont {L.}~\bibnamefont {Kahle}}, \bibinfo {author} {\bibfnamefont {R.}~\bibnamefont {Häuselmann}}, \bibinfo {author} {\bibfnamefont {D.}~\bibnamefont {Gresch}}, \bibinfo {author} {\bibfnamefont {T.}~\bibnamefont {Müller}}, \bibinfo {author} {\bibfnamefont {A.~V.}\ \bibnamefont {Yakutovich}}, \bibinfo {author} {\bibfnamefont {C.~W.}\ \bibnamefont {Andersen}}, \bibinfo {author} {\bibfnamefont {F.~F.}\ \bibnamefont {Ramirez}}, \bibinfo {author} {\bibfnamefont {C.~S.}\ \bibnamefont {Adorf}}, \bibinfo {author} {\bibfnamefont {F.}~\bibnamefont {Gargiulo}}, \bibinfo {author} {\bibfnamefont {S.}~\bibnamefont {Kumbhar}}, \bibinfo {author} {\bibfnamefont {E.}~\bibnamefont {Passaro}}, \bibinfo {author} {\bibfnamefont
  {C.}~\bibnamefont {Johnston}}, \bibinfo {author} {\bibfnamefont {A.}~\bibnamefont {Merkys}}, \bibinfo {author} {\bibfnamefont {A.}~\bibnamefont {Cepellotti}}, \bibinfo {author} {\bibfnamefont {N.}~\bibnamefont {Mounet}}, \bibinfo {author} {\bibfnamefont {N.}~\bibnamefont {Marzari}}, \bibinfo {author} {\bibfnamefont {B.}~\bibnamefont {Kozinsky}},\ and\ \bibinfo {author} {\bibfnamefont {G.}~\bibnamefont {Pizzi}},\ }\href {https://doi.org/10.1038/s41597-020-00638-4} {\bibfield  {journal} {\bibinfo  {journal} {Sci. Data}\ }\textbf {\bibinfo {volume} {7}},\ \bibinfo {pages} {300} (\bibinfo {year} {2020})}\BibitemShut {NoStop}%
\bibitem [{\citenamefont {Zsurka}\ \emph {et~al.}(2024)\citenamefont {Zsurka}, \citenamefont {Wang}, \citenamefont {Legendre}, \citenamefont {Miceli}, \citenamefont {Serra}, \citenamefont {Grützmacher}, \citenamefont {Schmidt}, \citenamefont {Rüßmann},\ and\ \citenamefont {Moors}}]{dataset}%
  \BibitemOpen
  \bibfield  {author} {\bibinfo {author} {\bibfnamefont {E.}~\bibnamefont {Zsurka}}, \bibinfo {author} {\bibfnamefont {C.}~\bibnamefont {Wang}}, \bibinfo {author} {\bibfnamefont {J.}~\bibnamefont {Legendre}}, \bibinfo {author} {\bibfnamefont {D.~D.}\ \bibnamefont {Miceli}}, \bibinfo {author} {\bibfnamefont {L.}~\bibnamefont {Serra}}, \bibinfo {author} {\bibfnamefont {D.}~\bibnamefont {Grützmacher}}, \bibinfo {author} {\bibfnamefont {T.~L.}\ \bibnamefont {Schmidt}}, \bibinfo {author} {\bibfnamefont {P.}~\bibnamefont {Rüßmann}},\ and\ \bibinfo {author} {\bibfnamefont {K.}~\bibnamefont {Moors}},\ }\bibfield  {journal} {\bibinfo  {journal} {Materials Cloud Archive}\ }\textbf {\bibinfo {volume} {2024.X}},\ \href {https://doi.org/10.24435/materialscloud:mx-bn} {10.24435/materialscloud:mx-bn} (\bibinfo {year} {2024})\BibitemShut {NoStop}%
\bibitem [{\citenamefont {Rüßmann}\ \emph {et~al.}(2023)\citenamefont {Rüßmann}, \citenamefont {{Antognini Silva}}, \citenamefont {Aliberti}, \citenamefont {Bröder}, \citenamefont {Janssen}, \citenamefont {Mozumder}, \citenamefont {Struckmann}, \citenamefont {Wasmer},\ and\ \citenamefont {Blügel}}]{aiida-kkr-code}%
  \BibitemOpen
  \bibfield  {author} {\bibinfo {author} {\bibfnamefont {P.}~\bibnamefont {Rüßmann}}, \bibinfo {author} {\bibfnamefont {D.}~\bibnamefont {{Antognini Silva}}}, \bibinfo {author} {\bibfnamefont {R.}~\bibnamefont {Aliberti}}, \bibinfo {author} {\bibfnamefont {J.}~\bibnamefont {Bröder}}, \bibinfo {author} {\bibfnamefont {H.}~\bibnamefont {Janssen}}, \bibinfo {author} {\bibfnamefont {R.}~\bibnamefont {Mozumder}}, \bibinfo {author} {\bibfnamefont {M.}~\bibnamefont {Struckmann}}, \bibinfo {author} {\bibfnamefont {J.}~\bibnamefont {Wasmer}},\ and\ \bibinfo {author} {\bibfnamefont {S.}~\bibnamefont {Blügel}},\ }\bibfield  {journal} {\bibinfo  {journal} {Zenodo}\ }\href {https://doi.org/10.5281/zenodo.3628251} {10.5281/zenodo.3628251} (\bibinfo {year} {2023})\BibitemShut {NoStop}%
\end{thebibliography}%

\end{document}